%% file: main.tex
\documentclass[a4paper,3p,11pt,sort&compress]{elsarticle}

\RequirePackage{snapshot}

\usepackage[Symbol]{upgreek}
\usepackage{mathptmx}
\usepackage[framed,numbered,autolinebreaks,useliterate]{mcode}
\usepackage[T1]{fontenc}
\usepackage[latin9]{inputenc}
\usepackage{amsmath}
\usepackage{amssymb}
\usepackage{graphicx}
\usepackage{setspace}
\usepackage{esint}
\PassOptionsToPackage{normalem}{ulem}
\usepackage{ulem}
\usepackage{marvosym}
\usepackage{multirow}
\usepackage{makecell}
\usepackage{soul} 
\usepackage{physics}
\usepackage{xfrac}
\usepackage{dsfont}
\usepackage[hidelinks]{hyperref}
\usepackage{nomencl}
\usepackage{adjustbox}
\usepackage{mathtools}
\usepackage{nicefrac}
\usepackage[hang,flushmargin]{footmisc}
\usepackage{amsthm}

\DeclareMathAlphabet\mathbfcal{OMS}{cmsy}{b}{n}

\makeatletter
\newcommand{\shorteq}{%
  \settowidth{\@tempdima}{-}
  \resizebox{\@tempdima}{\height}{=}%
}
\makeatother

\makeatletter
\renewcommand*\env@matrix[1][*\c@MaxMatrixCols c]{%
\hskip -\arraycolsep
\let\@ifnextchar\new@ifnextchar
\array{#1}}
\makeatother

\DeclareMathSymbol{\shortminus}{\mathbin}{AMSa}{"39}
\newcommand{\shortm}{\negthinspace\shortminus\negthinspace}

\soulregister\cite7
\soulregister\ref7
\soulregister\pageref7

\def\*#1{\mathbf{#1}}

\usepackage[bordercolor=white,backgroundcolor=gray!30,linecolor=black,colorinlistoftodos]{todonotes}

\usepackage{hyperref}

\makeatletter

\newcommand{\vt}[1]{\mathbf{#1}}                    
\newcommand{\vg}[1]{\bm{#1}}                        
\newcommand{\Her}[0]{\mathrm{H}}                      
\newcommand{\I}[0]{\mathrm{i}\mkern1mu}                      

\newcommand{\Ez}[0]{\mathbf{E}_0}

\newcommand{\Eo}[0]{\mathbf{E}_1}
\newcommand{\Et}[0]{\mathbf{E}_2}
\newcommand{\M}[0]{\mathbf{M}}

\newcommand*{\eten}{\*C} 								

\newcommand{\Mflow}[0]{\mathbf{L}}

\newcommand{\dotP}[0]{\scalebox{1}{\textbullet}}            

\newcommand{\egv}[0]{\vg{\upphi}}                  

\newcommand{\depa}[0]{\chi}                  
\newcommand{\norma}[0]{n}                  
\newcommand{\dlFreq}[0]{a_0}                  

\DeclareMathSymbol{\shortminus}{\mathbin}{AMSa}{"39}

\DeclareMathAlphabet\mathbfcal{OMS}{cmsy}{b}{n}      
\DeclareMathAlphabet\mathcal{OMS}{cmsy}{m}{n}      

\newcommand{\eq}{\begin{eqnarray}}
\newcommand{\nq}{\end{eqnarray}}
\newcommand{\dirvec}[1]{\mathbf{e}_{#1}}

\newcommand\ten[1]{\mathbf{#1}}

\usepackage{subfig} 
\usepackage{bm}
\usepackage{color}
\usepackage{epstopdf}

\newcommand{\tikzcircle}[2][red,fill=red]{\tikz[baseline=-0.5ex]\draw[#1,radius=#2] (0,0) circle ;}%

\makeatletter
\def\ps@pprintTitle{%
  \let\@oddhead\@empty
  \let\@evenhead\@empty
  \let\@oddfoot\@empty
  \let\@evenfoot\@oddfoot
}
\makeatother

\newcommand*\xbar[1]{%
  \hbox{%
    \vbox{%
      \hrule height 0.5pt 
      \kern0.3ex
      \hbox{%
        \kern-0.1em
        \ensuremath{#1}%
        \kern-0.1em
      }%
    }%
  }%
}

\newsavebox\myboxA
\newsavebox\myboxB
\newlength\mylenA

\newcommand*\xoverline[2][0.75]{%
    \sbox{\myboxA}{$\m@th#2$}%
    \setbox\myboxB\null
    \ht\myboxB=\ht\myboxA%
    \dp\myboxB=\dp\myboxA%
    \wd\myboxB=#1\wd\myboxA
    \sbox\myboxB{$\m@th\overline{\copy\myboxB}$}
    \setlength\mylenA{\the\wd\myboxA}
    \addtolength\mylenA{-\the\wd\myboxB}%
    \ifdim\wd\myboxB<\wd\myboxA%
       \rlap{\hskip 0.5\mylenA\usebox\myboxB}{\usebox\myboxA}%
    \else
        \hskip -0.5\mylenA\rlap{\usebox\myboxA}{\hskip 0.5\mylenA\usebox\myboxB}%
    \fi}

\usepackage{tikz,pgfplots,grffile} 
\pgfplotsset{compat=newest}
\usetikzlibrary{calc}

\input{myColorsMatlab.tex}

\journal{\ }

\newdefinition{rmk}{Remark}

\makeatother
\PassOptionsToPackage{numbers,sort&compress}{natbib}

\makenomenclature
\begin{document}
\makeatletter
\def\bm@pmb@#1{{%
      \setbox\tw@\hbox{$\m@th\mkern.25mu$}%
      \mathchoice
      \bm@pmb@@\displaystyle\@empty{#1}%
      \bm@pmb@@\textstyle\@empty{#1}%
      \bm@pmb@@\scriptstyle\defaultscriptratio{#1}%
      \bm@pmb@@\scriptscriptstyle\defaultscriptscriptratio{#1}}}
\makeatother

\title{Computing leaky waves in semi-analytical waveguide models\\ by exponential residual relaxation}

\author[ovgu]{Hauke~Gravenkamp\corref{cor1}}
\ead{hauke.gravenkamp@ovgu.de}

\author[lj]{Bor~Plestenjak}
\ead{bor.plestenjak@fmf.uni-lj.si}

\author[espci]{Daniel~A.~Kiefer}
\ead{daniel.kiefer@espci.fr}

\address[ovgu]{Institute of Materials, Technologies and Mechanics, Otto von Guericke University Magdeburg, 39106 Magdeburg, Germany}

\address[lj]{IMFM and Faculty of Mathematics and Physics, University of Ljubljana, Jadranska 19, SI-1000 Ljubljana, Slovenia}

\address[espci]{Institut Langevin, ESPCI Paris, Universit\'e PSL, 75005 Paris, France}

\cortext[cor1]{Corresponding author}

\begin{abstract}\noindent
Semi-analytical methods for the modeling of guided waves in structures of constant cross-section lead to frequency-dependent polynomial eigenvalue problems for the wavenumbers and mode shapes. Solving these eigenvalue problems for a range of frequencies results in continuous eigencurves that are of relevance in practical applications of ultrasonic measurement systems. Recent research has shown that eigencurves of parameter-dependent eigenvalue problems can alternatively be computed as solutions of a system of ordinary differential equations, which are obtained by postulating an exponentially decaying residual of a modal solution. This general concept for solving parameter-dependent matrix equations is, in this context, known as Zeroing Neural Networks or Zhang Neural Networks (ZNN).  We exploit this idea to develop an efficient method for computing the dispersion curves of plate structures coupled to unbounded solid or fluid media. In these scenarios, the alternative formulation is particularly useful since the boundary conditions give rise to nonlinear terms that severely hinder the application of traditional solvers.
\end{abstract}
\maketitle

\section{Introduction\label{sec:intro}}
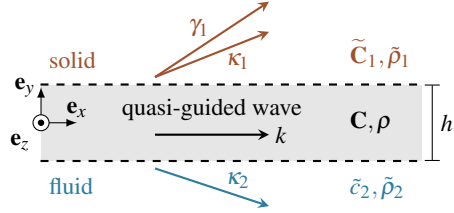
\begin{figure}\centering
  \subfloat{\input{plate_geometry_embedded}}
  \caption{Plate of thickness $h$, stiffness tensor $\eten$, and mass density $\rho$ in contact with unbounded media of different properties. For illustration, the plate is coupled both to another solid ($\widetilde{\ten{C}}_1, \tilde{\rho}_1$) and a fluid halfspace (wave velocity $\tilde{c}_2$, density $\tilde{\rho}_2$) at the top and bottom surface, respectively, while all combinations of fluid, solid, and free surface are generally possible. The wavenumber of a (quasi-)guided wave mode is denoted by $k$, while the free-field wavenumbers of longitudinal and shear waves in the unbounded media are $\kappa_{1,2}$ and $\gamma_1$, respectively.   \label{fig:plate_geometry_embedded}}
\end{figure}

\noindent
Elastic waves propagating along thin-walled structures, known as guided waves, occur in various engineering and scientific applications. In the ultrasonic range, they are widely used in non-destructive testing (NDT) and material characterization, while in a broader context, they also play a role in soil dynamics and earthquake engineering. A particularly interesting and relevant case arises when such waveguides are coupled to or immersed in an unbounded medium, whether solid or fluid, see Fig.~\ref{fig:plate_geometry_embedded}. Examples include layered soils \cite{Flitman1962,Kausel2022} and plate structures in NDT applications \cite{Drinkwater2003,Pelat2011,Pistone2015}.

Guided waves are characterized by a distinct modal behavior governed by the boundary conditions at the structure's surfaces. At any given frequency, an infinite number of modes exist; however, only a finite subset propagates with sufficiently low attenuation to be practically relevant. The computation of these modes' wavenumbers over a range of frequencies yields dispersion curves, which are essential in many applications for predicting wave propagation behavior. For simple cases like homogeneous plates with traction-free surfaces, implicit closed-form expressions for dispersion relations exist \cite{Lamb1917c,Kausel2013b}. More complex structures, such as layered plates or cylindrical waveguides, require more general formulations like the Transfer Matrix Method, Global Matrix Method, or Stiffness Matrix Method, together with mode-tracing techniques \cite{Knopoff1964,Nayfeh1991b}.
Over the past few decades, semi-analytical methods have emerged as the standard for most practical problems. These methods discretize the waveguide's cross-section using finite elements or similar techniques, giving rise to an eigenvalue formulation whose solutions yield wavenumbers and mode shapes.  Notable representatives include the semi-analytical finite element method (SAFE) \cite{Dong1972,Bartoli2006,Galan2002b}, the thin layer method (TLM) \cite{Kausel1977,Kausel1986a,Kausel2004}, and a variant of the scaled boundary finite element method \cite{Gravenkamp2012,Krome2017}, which are all widely adopted due to their robustness and ability to capture all solutions that the chosen discretization can represent.
Despite their advantages, semi-analytical methods face challenges in describing waveguides that are in contact with unbounded media. Approaches such as perfectly matched layers (PML) \cite{Basu2004}, boundary element methods \cite{Mazzotti2013b}, and approximate techniques using dashpots \cite{Gravenkamp2014b, Gravenkamp2015} have been developed to address this issue. On the other hand, incorporating exact boundary conditions typically leads to nonlinear eigenvalue problems \cite{Gravenkamp2014c,Gravenkamp2025}, which are notoriously difficult to solve \cite{Guttel2017,Jarlebring:2017:TIAR,Mehrmann:2004:NLEVP}. In special cases, such as homogeneous fluid loading, the problem can be linearized and efficiently solved \cite{Kiefer2019}. Iterative solutions have also been proposed for fluid loading scenarios \cite{Gravenkamp2014c}.

Recently, we demonstrated that the nonlinear eigenvalue problem arising in more general cases, such as layered plate structures coupled to fluid and/or solid halfspaces, can be stated as a multiparameter eigenvalue problem \cite{Gravenkamp2025}. This reformulation allows for the application of recently developed solution techniques \cite{HKP_JD2EP,MP_Q2EP}, implemented in the Matlab toolbox MultiParEig \cite{multipareig_2023}. While this approach is highly robust and capable of finding all solutions, it is computationally demanding. The cost increases rapidly with the size of the finite element matrices and the number of additional parameters, restricting its application to relatively small cases.

In this paper, we explore an alternative approach to computing eigencurves based on a methodology that was presented by Zhang et al.\ \cite{Zhang2001a,Zhang2002a, Zhang2005} and further developed by Uhlig et al.\ \cite{Uhlig2019, Uhlig2020a, Uhlig2024} for the solution of different parameter-dependent matrix equations. The key idea is to transform the matrix equation (in our application, a parameter-dependent nonlinear eigenvalue problem) into an ordinary differential equation (ODE) by assuming that the solution's residual decays exponentially with respect to the free parameter. The resulting ODE can then be solved using standard numerical techniques, such as Runge-Kutta schemes. This concept arose in this form in the study of recurrent neural networks (see also, e.g., \cite{Cichocki1992, Wang1993}) and is hence known to many as Zeroing Neural Networks \cite{Jin2017, Wang2024} or Zhang Neural Networks \cite{Sun2016,Uhlig2024}, both abbreviated as ZNN. In the context of finite element models, the same principal idea has been developed to enforce constraints in transient nonlinear problems \cite{Gravenkamp2023a}. There are also several somewhat similar methodologies such as dynamic relaxation \cite{Otter1965}, artificial compressibility \cite{Chorin1967, Madsen2006}, or Baumgarte stabilization \cite{Baumgarte1972}. 

In \cite{Gravenkamp2022}, we have briefly considered this approach for computing eigencurves of the free waveguide problem but stated that we were `not yet sure about this approach's practical usefulness.' The free waveguide problem is generally easy to linearize and efficiently solvable by standard methods, making an alternative mode-tracing approach less relevant. However, we now recognize that this technique is particularly advantageous for leaky waveguides, as it can directly incorporate the involved nonlinearities. This allows computations to be performed on the original matrices, in contrast to the transformation into a multiparameter eigenvalue problem, which ultimately requires the solution of a much larger linear problem.
Furthermore, in the case of leaky waves, the focus is typically on a small fraction of the modes---specifically those that radiate energy away from the waveguide and exhibit relatively low attenuation. A core advantage of the proposed approach is that it follows individual modes during computation, unlike traditional semi-analytical solution procedures that compute eigenvalues at predefined frequency or wavenumber points without keeping track of their evolution. This bears some conceptual resemblance to mode-tracing techniques used in the Global Matrix Method and other analytical approaches, which often employ extrapolation of known solutions necessary to establish starting values for root-finding algorithms. 

This paper focuses on the development of an algorithm for computing eigencurves of the nonlinear eigenvalue problem describing leaky waves in plate structures. For brevity, we will not provide details on the derivation of said eigenvalue problem and instead refer to some of the many papers that contain the complete formulation. We will pose the known problem directly in Section~\ref{sec:problem} and then explain the fundamental concept of deriving a system of ODEs from a given objective function in Section~\ref{sec:formulation}. We then apply this formulation to the problem at hand, i.e., the nonlinear eigenvalue problem, in Section~\ref{sec:specificCases} and discuss the issue of efficiently obtaining good initial values to start the mode-tracing in Section~\ref{sec:initial}. Finally, we present a selection of numerical examples of increasing complexity in Section~\ref{sec:numex} before drawing conclusions from our findings in Section~\ref{sec:conclusion}.

\section{Problem statement\label{sec:problem}} \noindent
The semi-discrete version of the guided wave formulation, when the coupling to unbounded fluid or solid media
is included, leads to a nonlinear eigenvalue problem for wavenumbers $k$ and eigenvectors $\egv$ with the frequency $\omega$ playing the role of the continuous parameter. The eigenvector includes the nodal displacements arising from the finite-element discretization of the plate's cross-section, as well as the amplitudes of the displacements or pressure describing the unbounded domains at the plate's surfaces. The detailed derivation of the formulation in the version used here is presented in \cite{Gravenkamp2025} and is not repeated here. Further details on the underlying semi-analytical model for general waveguides can be found in, e.g., \cite{Gravenkamp2012, Gravenkamp2014c} and many others. The finite-element matrices used in the following were computed with the help of the open-source Matlab implementation \textit{SAMWISE} \cite{Gravenkamp2024a}.
Since the form of the waveguide problem discussed here involves only terms in $\omega^2$, we introduce $\mu = \omega^2$ and write
\begin{equation}
  \left(-k^2\, \Ez + \I k\,\Eo-\Et+\mu\,\M+\*R(k,\mu)\right)\egv=\vt{0}. \label{eq:GWproblem_embedded}
\end{equation}
The term $\*R(k,\mu)$ incorporates the nonlinearities that arise due to coupling to the unbounded halfspaces at the top and bottom of the plate structure; it reads
\begin{equation}\label{ex:nonpoly1}
  \*R(k,\mu) = \sum\limits_{j=1}^6 b_j\, \xi_j\, \*R_j
\end{equation}
with 
\begin{equation}
  b_j = \begin{cases}
    \I & j \in \mathrlap{\{1,2\}} \hphantom{\{3,4,5,6\}} \qquad \text{fluid halfspace},\\
    k & j \in \{3,4,5,6\} \qquad \text{solid halfspace},
  \end{cases}
\end{equation}
Here, the symbol $\xi$ represents vertical wavenumbers of partial waves in the unbounded domains, and $\I$ is the imaginary unit.
The sparse coupling matrices $\*R_j$ have been derived in detail previously \cite{Gravenkamp2025}. For simplicity in notation, we always include all six terms (one/two partial waves in each adjacent fluid/solid), and for any halfspace that is not present in the current model, we set the corresponding $\*R_j$ matrices to $\*0$ and $\xi_j$ to an arbitrary finite value. Hence, we introduce here the parameter $b_j$ to distinguish the slightly different terms for solid and fluid media. This notation implies that $j \in \{1,2\}$ always correspond to variables associated with the acoustic fluids (if present), while $j \in \{3,4,5,6\}$ are reserved for solid media.
The vertical wavenumber $\xi_j$ is related to the corresponding free-field wave speed $c_j$ (for every halfspace present in the model) via
\begin{equation}\label{eq:xi}
  \xi_{j}(k,\mu) = \pm\sqrt{\frac{\mu}{c_j^2} - k^2}.
\end{equation}
Note that, in the above relationship, the vertical wavenumber depends on the eigenvalue $k$ in a nonlinear fashion, which is
the main challenge in solving this class of problems.

\section{Solution approach\label{sec:formulation}}\noindent
\subsection{General concept}\noindent
The basic idea, as outlined in \cite{Zhang2002a, Zhang2005, Uhlig2019, Uhlig2020a, Uhlig2024} and applied to the free waveguide problem in \cite{Gravenkamp2022}, is the following. Assume we want to minimize a parameterized objective function $f(y(t),t)$, i.e., solve
\begin{equation} \label{eq:f}
  f(y(t),t) = 0
\end{equation}
with a given parameter $t\in[t_0,t_1]$. Thus, the goal is to find a function $y(t)$, such that the objective function vanishes for all $t$ within the given interval.
To address this problem numerically, we postulate the existence of the following derivative
\begin{equation}\label{eq:fODE}
  f'(y(t),t) = -\depa f(y(t),t),
\end{equation}
which implies an exponentially decreasing residual of \eqref{eq:f}. Here, the prime symbol denotes the total derivative with respect to $t$, and $\depa$ is an algorithmic constant to be chosen later.
Equation~\eqref{eq:fODE} represents an ordinary differential equation (ODE) that can be solved numerically, provided that $y$ is known at some initial value, say $y(t_0) \eqqcolon y_0$.
Hence, instead of Eq.~\eqref{eq:f}, we will solve the initial value problem:\\

\noindent \textit{Let $f(y(t),t)$ be a differentiable objective function and $\mathcal{I} = (t_0, t_1)$ an open interval. Find $y(t)$ such that}
\begin{subequations}
  \begin{align}
    f'(y(t),t) &= -\depa f(y(t),t),\quad t\in\mathcal{I}\\
    y(t_0)  &= y_0.
  \end{align}
\end{subequations}

\subsection{Introductory example\label{eq:minimalEx}}
\begin{figure}
  \subfloat[]{\includegraphics[width=0.5\textwidth]{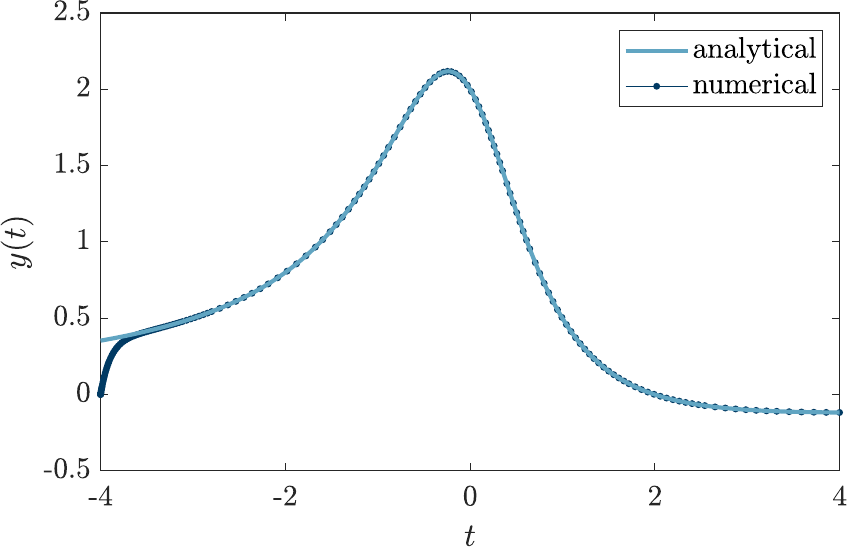}}
  \subfloat[]{\includegraphics[width=0.5\textwidth]{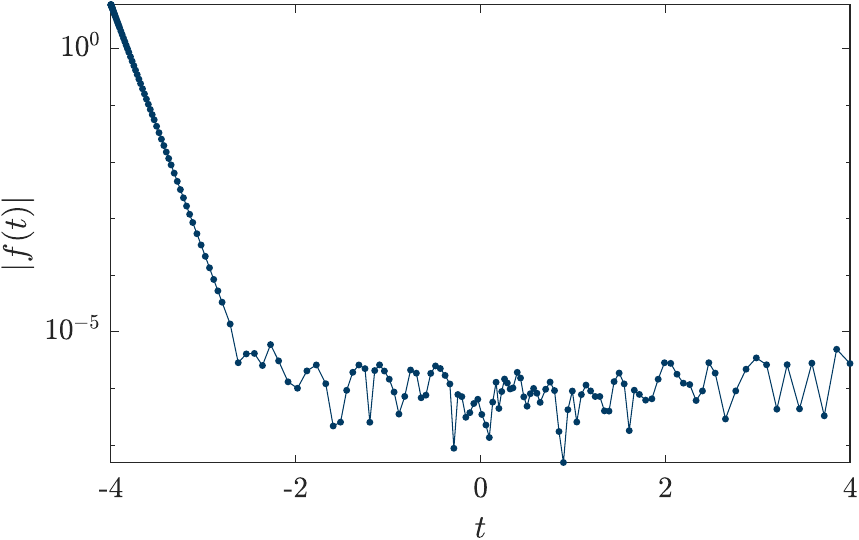}}
  \caption{Analytical and numerical solution of the introductory example (a) and residual of the numerical solution (b). \label{fig:minimal_results}}
\end{figure}

\noindent
Let us consider the objective function
\begin{equation}\label{eq:f_minimal}
  f(y(t),t) = t + (t^2 + 1)y -2
\end{equation}
with the derivative
\begin{equation}\label{eq:df_minimal}
  f'(y(t),t) = 1 + 2ty + (t^2 + 1)y'.
\end{equation}
This problem has a simple analytical solution that we pose here for verifying the results; namely, the objective function vanishes for
\begin{equation}
  y_\mathrm{e}(t) = \frac{2-t}{t^2+1},
\end{equation}
i.e.,
\begin{equation}
  f(y_\mathrm{e}(t),t) = 0\quad \forall\ t.
\end{equation}
To demonstrate the numerical procedure, we substitute \eqref{eq:f_minimal} and \eqref{eq:df_minimal} into \eqref{eq:fODE} and obtain
\begin{equation}\label{eq:fODE_minimal}
  (t^2 + 1)\,y' = -\depa \big(t + (t^2 + 1)y -2 \big) -1-2t y.
\end{equation}
The ODE \eqref{eq:fODE_minimal} can be integrated numerically starting from some known value of $y$. Here, we employ a Runge-Kutta based solver with a variable stepsize as implemented in Matlab's function \emph{ode15s}.
For a more efficient evaluation, we can provide the solver with the Jacobian, which is obtained as the derivative of the right-hand side of \eqref{eq:fODE_minimal} with respect to $y$:
\begin{equation}\label{eq:Jac_minimal}
  J(t) = -\depa (t^2 + 1)-2t.
\end{equation}
Say we want to solve Eq.~\eqref{eq:f_minimal} within an interval $t\in[-4,4]$.
As a starting value, we could, of course, use the known exact solution at a given $t$-value. However, to demonstrate the robustness of this approach, particularly in cases where an exact solution is unknown, we choose an approximate initial value. Noting that
\begin{equation}
  \lim_{t\to\pm\infty} \ y_\mathrm{e}(t) = 0, 
\end{equation}
we approximate
\begin{equation}
  y_\mathrm{e}(-4) \approx 0
\end{equation}
and, hence, select $t_0 = -4$, $y_0 = 0$. Furthermore, we choose $\depa = 10$ and set the relative tolerance of the numerical solver to $10^{-6}$.
Results of the computed solution and the corresponding residual are shown in Fig.~\ref{fig:minimal_results}. It is worth noting that, despite the large error of the initial condition (the exact value is $y_\mathrm{e}(-4) = 6/17 \approx 0.35$), the numerical result rapidly approaches the exact solution and ultimately yields residuals below $10^{-5}$.
In fact, the algorithm converges even if the initial condition is off by several orders of magnitude.

\subsection{Extension to coupled systems of equations\label{sec:coupled}}\noindent
We observe that the procedure outlined above is adjusted straightforwardly to treat systems that depend on several functions of the parameter $t$. Namely, the minimization of a system of objective functions
\begin{equation}
  \*f(\*y(t), t) \coloneqq 
  \begin{bmatrix}
    f_1(y_1(t),y_2(t),...,y_n(t),t)\\
    f_2(y_1(t),y_2(t),...,y_n(t),t)\\
    \vdots \\
    f_n(y_1(t),y_2(t),...,y_n(t),t)
  \end{bmatrix}
  = \*0
\end{equation}
is posed as the solution of the initial value problem
\begin{subequations}
  \begin{align}\label{eq:ODE_multi}
    \*f'(\*y(t), t) &= -\depa\, \*f(\*y(t), t),\quad t\in\mathcal{I}\\
    \*y(t_0)  &= \*y_0.
  \end{align}
\end{subequations}
Here, we have assumed that the same algorithmic constant $\depa$ is chosen for all entries of $\*f(\*y(t), t)$. Otherwise, $\depa$ is to be replaced by a corresponding  diagonal matrix $\*X = \operatorname{diag}(\depa_1, \depa_2, ..., \depa_n)$.

\subsection{Application to eigenvalues of matrix functions\label{sec:eigencurves}}\noindent
We will now apply the same technique to compute the eigencurves of a matrix function, i.e., to solve an eigenvalue problem that depends on one parameter:
\begin{equation}
  \Mflow(k,\mu)\,\egv=\vt{0}. \label{eq:NLEVP_matrixFlow}
\end{equation}
The specific case of (leaky) guided waves will be addressed in the following section.
For easier comparison with the previous literature, the eigenvalue and eigenvector are denoted as $k = k(\mu)$, $\egv = \egv(\mu)$, and $\mu = \omega^2$ is the free parameter.
As the eigenvectors are only defined up to a multiplicative constant, we need to choose a normalization in order to obtain unique results, e.g.,
\begin{equation}\label{eq:normalization}
  \norma(\egv) \coloneqq \egv^\Her \egv -1 =0.
\end{equation}
Hence, we define the objective function
\begin{equation}
  \*f(\egv,k,\mu) =
  \begin{bmatrix}
    \Mflow(k,\mu)\,\egv \\
    \norma(\egv)
  \end{bmatrix} \label{eq:evp_objective}
\end{equation}
and write the total derivative with respect to the parameter $\mu$ as
\begin{equation}
  \*f'(\egv,k,\mu) =
  \begin{bmatrix}
    \Mflow(k,\mu)\egv'
    +  \Mflow'(k,\mu)\egv \\
    \vg{n}_{\egv}(\egv)\, \egv'
  \end{bmatrix}. \label{eq:leaky_Fdomega}
\end{equation}
\textit{Remark: The normalization defined by Eq.~\eqref{eq:normalization} is not complex differentiable with respect to $\egv$, though it is differentiable with respect to the real and imaginary parts $\mathfrak{R}(\egv)$, $\mathfrak{I}(\egv)$, and the Wirtinger derivatives exist \cite{Amin2011}, permitting a well-defined complex gradient \cite{Brandwood1983, VanDenBos1994}. It is known in the context of optimization problems that we can employ the approximation  $2\egv^\Her$ (analogously to real-valued vectors) to obtain a linearization, e.g., in Newton's method \cite{Lu2022}. Here, we will use the same concept to approximate the derivative with respect to $\egv$ that we require in the following steps. There are other possible normalizations; in particular, a popular choice is to use $n(\egv) = \egv_\mathrm{c}^\Her \egv - 1$, with some constant vector $\egv_\mathrm{c}$ (e.g., the initial value). While this approach generally works and avoids the theoretical issue of a non-differentiable normalization, we found that the one defined by Eq.~\eqref{eq:normalization} resulted in a slightly more robust computation in our examples. Hence, we will stick with this version and the general notation $\vg{n}_{\egv}$, irrespective of the normalization we employ, keeping in mind that, in the case of Eq.~\eqref{eq:normalization}, $\vg{n}_{\egv}(\egv) = 2\egv^\Her$ is not a true complex derivative.
}\\

\noindent
We continue by isolating terms in $\*f'(\egv,k,\mu) $ that are multiplied by $\egv$, $\egv'$, or $k'$. To this end,
the total derivative $\Mflow'(k,\mu)$ is written using the partial derivatives as
\begin{equation}\label{eq:totalDerivative}
  \Mflow'(k,\mu) = \frac{\mathrm{d}}{\mathrm{d}\mu} \Mflow(k,\mu) = 
  \frac{\partial}{\partial\mu} \Mflow(k,\mu) + \frac{\partial}{\partial k} \Mflow(k,\mu)\,\frac{\mathrm{d}}{\mathrm{d}\mu}k,
\end{equation}
which we abbreviate as
\begin{equation}\label{eq:totalDerivative2}
  \Mflow'(k,\mu) \eqqcolon
  \Mflow_\mu(k,\mu) + \Mflow_k(k,\mu)\,k',
\end{equation}
such that
\begin{equation}\label{eq:evp_fdmu}
  \*f'(\egv,k,\mu) =
  \begin{bmatrix}
    \Mflow(k,\mu)\,\egv' + \Mflow_k(k,\mu)\,\egv\,k'  + \Mflow_\mu(k,\mu)\,\egv \\
    \vg{n}_{\egv}(\egv)\, \egv'
  \end{bmatrix}. 
\end{equation}
The precise definition of these terms will be specified later for the cases of interest. 
Substituting Eqs.~\eqref{eq:evp_objective} and \eqref{eq:evp_fdmu} into \eqref{eq:ODE_multi}, we find that the system of ODEs is of the form
\begin{equation}
  \begin{bmatrix}
    \Mflow(k,\mu) & \Mflow_k(k,\mu)\,\egv \\
    \vg{n}_{\egv}(\egv)      & 0
  \end{bmatrix}
  \begin{bmatrix}
    \egv' \\
    k'
  \end{bmatrix}
  =
  - \begin{bmatrix}
    \depa_1\, \Mflow(k,\mu)\,\egv \\
    \depa_2\, \norma(\egv)
  \end{bmatrix}
  -
  \begin{bmatrix}
    \Mflow_\mu(k,\mu)\,\egv \\
    0
  \end{bmatrix}
  , \label{eq:leaky_ODE}
\end{equation}
which we will abbreviate for later use as
\begin{equation}
  \*{A}(\egv,k,\mu) \begin{bmatrix}
    \egv' \\
    k'
  \end{bmatrix}
  = \*{b}(\egv,k,\mu).
\end{equation}
Note that we use two different decay parameters $\depa_1$, $\depa_2$ to account for a potentially different behavior of the matrix equation \eqref{eq:NLEVP_matrixFlow} and the normalization \eqref{eq:normalization}.
The Jacobian is given as
\begin{equation}
  \*J(\egv,k,\mu) =
  \begin{bmatrix}
    \partial_{\egv} \*{b} & \partial_k \*{b}
  \end{bmatrix} =
  -\begin{bmatrix}
    \depa_1\, \Mflow(k,\mu) +  \Mflow_\mu(k,\mu) &
    \depa_1\, \Mflow_k(k,\mu)\,\egv +  \Mflow_{\mu k}(k,\mu)\,\egv \\
    \depa_2\, \vg{n}_{\egv}(\egv)                      &
    0
  \end{bmatrix}
  , \label{eq:leaky_Jac}
\end{equation}
with
\begin{equation}
  \Mflow_{\mu k} = \frac{\partial^2 }{\partial k \partial \mu} \Mflow(k,\mu) \,.
\end{equation}
Both $\egv$ and $k$ are generally complex-valued. It is interesting to note that \eqref{eq:leaky_ODE}  can be split into real and imaginary parts in order to optimize both simultaneously. Defining $\egv = \egv_r + \I\egv_i$ and $k = k_r + \I k_i$, we can rewrite \eqref{eq:leaky_ODE} as
\begin{equation}
  \begin{bmatrix}
    \mathfrak{R}(\Mflow) & -\mathfrak{I}(\Mflow)           & \mathfrak{R}(\Mflow_k\,\egv) & -\mathfrak{I}(\Mflow_k\,\egv)           \\
    \mathfrak{I}(\Mflow) & \phantom{-}\mathfrak{R}(\Mflow) & \mathfrak{I}(\Mflow_k\,\egv) & \phantom{-}\mathfrak{R}(\Mflow_k\,\egv) \\
    \mathfrak{R}(\vg{n}_{\egv})          & -\mathfrak{I}(\vg{n}_{\egv})                     & 0                      & 0                                 \\
    \mathfrak{I}(\vg{n}_{\egv})           & \phantom{-}\mathfrak{R}(\vg{n}_{\egv})         & 0                      & 0                                 \\
  \end{bmatrix}
  \begin{bmatrix}
    \egv_r' \\
    \egv_i' \\
    k_r'    \\
    k_i'    \\
  \end{bmatrix}
  =
  -\begin{bmatrix}
    \depa_1\, \mathfrak{R}(\Mflow\,\egv)    \\
    \depa_1\, \mathfrak{I}(\Mflow\,\egv)    \\
    \depa_2\, \mathfrak{R}(\norma) \\
    \depa_2\, \mathfrak{I}(\norma)
  \end{bmatrix}
  -
  \begin{bmatrix}
    \mathfrak{R}(\Mflow_\mu\,\egv) \\
    \mathfrak{I}(\Mflow_\mu\,\egv) \\
    0                         \\
    0
  \end{bmatrix}
  . \label{eq:leaky_ODE_split}
\end{equation}
While some implementations of solvers for systems of ODEs may recommend or even require this real-valued formulation, we will stick with the complex version in what follows.

\subsection{Incorporating algebraic constraints\label{sec:constraint}}\noindent
It is generally straightforward to formulate the approach for an arbitrary number of coupled equations, as mentioned in Section~\ref{sec:coupled}. Here, we only showcase one specific case that will be of particular importance to our current application, namely the inclusion of additional scalar equations. Assume the eigenvalue problem depends on another parameter $\xi = \xi(\mu)$, such that
\begin{equation}
  \Mflow(k,\xi,\mu)\,\egv=\vt{0}. \label{eq:NLEVP_matrixFlow_xi}
\end{equation}
In our case, the solution to \eqref{eq:NLEVP_matrixFlow_xi} satisfies a scalar equation
\begin{equation}
  g(k,\xi,\mu) = 0.
\end{equation}
Thus, Eq.~\eqref{eq:leaky_ODE} is extended as
\begin{equation}
  \begin{bmatrix}
    \Mflow(k,\xi,\mu) & \Mflow_\xi(k,\xi,\mu)\,\egv & \Mflow_k(k,\xi,\mu)\,\egv \\
    \*0             & g_\xi(k,\xi,\mu)           & g_k(k,\xi,\mu)   \\
    \vg{n}_{\egv}(\egv)      & 0 & 0
  \end{bmatrix}
  \begin{bmatrix}
    \egv' \\
    \xi' \\
    k'
  \end{bmatrix}
  =
  - \begin{bmatrix}
    \depa_1\, \Mflow(k,\xi,\mu)\,\egv \\
    \depa_1\, g(k,\xi,\mu)  \\
    \depa_2\, \norma(\egv)
  \end{bmatrix}
  -
  \begin{bmatrix}
    \Mflow_\mu(k,\xi,\mu)\,\egv \\
    g_\mu(k,\xi,\mu) \\
    0
  \end{bmatrix}
  , \label{eq:leaky_ODE_constraint}
\end{equation}
where the matrix entries are defined via 
\begin{subequations}
  \begin{align}
  \Mflow'(k,\xi,\mu) &\eqqcolon
  \Mflow_\mu(k,\xi,\mu)+ \Mflow_\xi(k,\xi,\mu)\,\xi' + \Mflow_k(k,\xi,\mu)\,k', \\
  g'(k,\xi,\mu) &\eqqcolon g_\mu(k,\xi,\mu) + g_\xi(k,\xi,\mu)\,\xi' + g_k(k,\xi,\mu)\,k',
\end{align}
\end{subequations}
and the Jacobian is extended correspondingly as
\begin{equation}
  \*J(\egv,k,\xi,\mu) =
  \begin{bmatrix}
    \partial_{\egv} \*{b}  & \partial_\xi \*{b} & \partial_k \*{b}
  \end{bmatrix} =
  -\begin{bmatrix}
    \depa_1\, \Mflow +  \Mflow_\mu &
    \depa_1\, \Mflow_\xi\,\egv + \Mflow_{\mu \xi}\,\egv&
    \depa_1\, \Mflow_k\,\egv + \Mflow_{\mu k}\,\egv \\
    \*0 
    & \depa_1\,  g_\xi + g_{\mu \xi}
    & \depa_1\,  g_k + g_{\mu k}\\
    \depa_2\, \vg{n}_{\egv}                     &
    0 & 0
  \end{bmatrix}
  . \label{eq:leaky_Jac_constrained}
\end{equation}
Several constraints of this form can be included analogously.

\newpage
\section{Specific case of (leaky or trapped) guided waves\label{sec:specificCases}}\noindent
\subsection{Overview}\noindent
We can now address the problem at hand, i.e., the nonlinear parameterized eigenvalue problem~\eqref{eq:GWproblem_embedded}. For clarity, we substitute
Eqs.~\eqref{ex:nonpoly1} and \eqref{eq:xi} into \eqref{eq:GWproblem_embedded} to appreciate the overall structure of the eigenvalue problem:
\begin{equation}
  \Big(-k^2\, \Ez + \I k\,\Eo-\Et+\mu\,\M + 
  \sum \limits_{j=1}^6 \pm b_j\,\sqrt{\mu/c_j^2 - k^2}\, \*R_j\Big)\egv=\vt{0}. \label{eq:GWproblem_embedded_combined}
\end{equation}
To apply the solution procedure outlined in Section~\ref{sec:eigencurves}, we only need to derive the terms 
$\Mflow_\mu(k,\mu)$, $\Mflow_k(k,\mu)$, $\Mflow_{\mu k}(k,\mu)$. As they are trivial to obtain, we will provide the relevant expressions without much further explanation. In the following, we will refer to this variant of the algorithm as \textit{\sc version~i}. We will see that this approach is remarkably robust and accurate for the most part -- with one important exception: The square-root function $\sqrt{z}$ is not complex differentiable for $\mathfrak{R}(z)\leq 0, \mathfrak{I}(z) = 0$ (and not even continuous at $\mathfrak{I}(z) = 0$ for any $\mathfrak{R}(z)< 0$).\footnote{This explanation follows the convention that $\sqrt{z}$ denotes the principal value, hence, its real part is chosen to be nonnegative. It is possible to use other definitions that would instead lead to a discontinuity, e.g., for positive imaginary numbers, resulting in the analogous challenge for different modes.}

While the vast majority of solutions exhibit a significant imaginary part of the wavenumber (i.e., attenuation due to leakage into the adjacent halfspace or material damping), some \textit{trapped modes} may be present that are characterized by a vanishing imaginary part. Famously, the \textit{quasi-Scholte modes} that exist in a solid plate immersed in an acoustic fluid propagate with a real wavenumber, say $k_S$, slightly larger than the free-field wavenumber in the fluid, and are thus characterized by 
\begin{equation}
  (\kappa^f)^2-k_S^2 < 0, \quad |(\kappa^f)^2-k_S^2| \ll 1, \quad \mathfrak{I}\big((\kappa^f)^2-k_S^2\big) = 0.
\end{equation}
Numerically, the consequence of this discontinuity in the coupling term is that the solution of the ODE is prone to strong oscillations when the argument of the square root approaches the negative real axis. The nonlinear solver essentially extrapolates based on previous solutions of the eigenvector and eigenvalue and performs iterations to minimize the residual of the nonlinear objective function. Hence, near the negative real axis, it becomes likely that, during an extrapolation or iteration, an approximation for the eigenvalue is found with the incorrect sign. Once this happens, the nonlinear iterations are unlikely to converge due to the discontinuity. 
Depending on the employed solver, this issue manifests either as erroneous solutions or a failure to adjust the stepsize within the acceptable error tolerance. 

To eliminate this issue of a discontinuous objective function, we consider an alternative variant of the proposed procedure that consists of treating the vertical wavenumbers $\xi_j$ as additional parameters and solving the coupled system of equations
\begin{subequations}
  \begin{align}
    \Big(-k^2\, \Ez + \I k\,\Eo-\Et+\mu\,\M + \sum\limits_{j=1}^6 b_j\,\xi_{j}\, \*R_j\Big)\egv &= \vt{0},\\
    \xi_1^2 - \frac{\mu}{c_1^2} + k^2 &= 0,\\
    \xi_2^2 - \frac{\mu}{c_2^2} + k^2 &= 0,\\\nonumber
    \vdots& \quad \label{eq:GWproblem_embedded_xiSquared}
  \end{align}
  \end{subequations}
Thus, for each partial wave in one of the unbounded domains (if present), we add a constraint equation in the sense of Section~\ref{sec:constraint}. Again, this formulation is straightforwardly obtained after deriving the terms in Eq.~\eqref{eq:leaky_ODE_constraint}, which will be given in Section~\ref{sec:versionTwo}. We will refer to this formulation as {\sc version ii}. Clearly, the advantage of this approach lies in the fact that the constraint equations are continuously differentiable with respect to all parameters $(k,\xi_j,\mu)$, hence, numerical instabilities near real-valued solutions are avoided.  However, this version can, in some cases, cause a different problem: As we introduce a constraint on $\xi_j^2$ rather than $\xi_j$, instabilities can occur if two solutions corresponding to the branches $+\sqrt{\mu/c_j^2 - k^2}$ and $-\sqrt{\mu/c_j^2 - k^2}$ are very similar. So far, we have encountered this issue only once (Example IV in Section~\ref{sec:numex_leaky}) in the case of rather extreme attenuation due to a small acoustic mismatch between the plate structure and the unbounded media. Nevertheless, it is worthwhile combining both versions into one highly robust algorithm. We propose using {\sc version i} as the default variant and switching to {\sc version ii} only when the argument of the square root function for any of the partial waves gets close to the negative real axis. Specifically, we use {\sc version ii} if $|\mathfrak{I}(\xi_j^2)|<0.01|\mathfrak{R}(\xi_j^2)|$ or $|\xi_j^2|<0.01$ for any of the partial waves.

\subsection{\sc version i\label{sec:versionOne}}\noindent
Let us now list the expressions obtained by applying the proposed approach to the two versions of the problem statement mentioned above.
The first version is based on Eq.~\eqref{eq:leaky_ODE} with 
\begin{equation}
  \Mflow(k,\mu) = -k^2\, \Ez + \I k\,\Eo-\Et+\mu\,\M+\*R(k,\mu)
\end{equation}
and
\begin{equation}
  \*R(k,\mu) = \sum\limits_{j=1}^6 b_j\, \xi_{j}\, \*R_j.
\end{equation}
Noting that 
\begin{equation}
  \frac{\partial}{\partial k}\xi_{j}(k,\mu) = -\frac{ k}{\xi_{j}}, \qquad \frac{\partial}{\partial \mu}\xi_{j}(k,\mu)= \frac{ 1}{2c_j^2\xi_{j}},
\end{equation}
and 
\begin{equation}
   b_{j,k} \coloneqq \frac{\mathrm{d}}{\mathrm{d} k} b_j =  \begin{cases}
    0 & j \in \mathrlap{\{1,2\}} \hphantom{\{3,4,5,6\}} \qquad \text{fluid halfspace,}\\
    1 & j \in \{3,4,5,6\} \qquad \text{solid halfspace,}
  \end{cases}
\end{equation}
the expressions in Eq.~\eqref{eq:leaky_ODE} are readily obtained as
\begin{align}
  \Mflow_k(k,\mu) =  \frac{\partial}{\partial k}\Mflow(k,\mu) &=
  -2k\, \Ez + \I \,\Eo
  + \sum\limits_{j=1}^6
  \left(b_{j,k}\,\xi_{j} -\frac{b_j\,k}{\xi_{j}}\right)\*R_j, \\
  \Mflow_\mu(k,\mu) =  \frac{\partial}{\partial \mu}\Mflow(k,\mu) &= \M
  + \sum\limits_{j=1}^6 \frac{b_j }{2\xi_{j}c_j^2}\, \*R_j.
\end{align}
For computing the Jacobian, we additionally require
\begin{align}
  \Mflow_{\mu k}(k,\mu) = \frac{\partial^2}{\partial k \partial \mu}\Mflow(k,\mu) &=
  \sum\limits_{j=1}^6
  \Big(
  \frac{b_{j,k}}{2\xi_{j}c_j^2}
  + \frac{b_j\,k}{2\xi_j^3 c_j^2}
  \Big) \, \*R_j.
\end{align}

\subsection{\sc version ii\label{sec:versionTwo}}\noindent
The second variant differs from the first one in that we treat the $\xi_j$ as additional parameters, such that
\begin{equation}
  \Mflow = \Mflow(k,\vg{\xi},\mu), \quad \*R = \*R(k,\vg{\xi},\mu)
\end{equation}
and include the constraint equations of the form 
\begin{equation}
  g_j(k,\xi_j,\mu) = \xi^2_j - \frac{\mu}{c_j^2} + k^2 = 0.
\end{equation}
The terms in Eq.~\eqref{eq:leaky_ODE_constraint} are thus obtained as
\begin{align}
  & \Mflow_\mu = \M, \\
  & \Mflow_k(k,\vg{\xi}) =
  -2k\, \Ez + \I \,\Eo + 
  \sum\limits_{j=1}^6 b_{j,k}\, \xi_j\*R_j, \\
  & \Mflow_\xi(k,\vg{\xi}) =
  \sum\limits_{j=1}^6 b_j\,\*R_j, \\
  & g_{j\mu}(k,\xi_j,\mu)
   = - \tfrac{1}{c_j^2},\\
  & g_{jk}(k,\xi_j,\mu)
   =  2k \\
  & g_{j\xi}(k,\xi_j,\mu)
  = 2\xi_j.
\end{align}
Computing the Jacobian via Eq.~\eqref{eq:leaky_Jac_constrained} is particularly simple, since
\begin{align}
  & g_{j\mu k} = g_{j\mu\xi}
  = 0,\\
  & \Mflow_{\mu k} = \Mflow_{\mu \xi} = \*0.
\end{align}

\subsection{On the choice of the decay parameters $\depa_1$, $\depa_2$ \label{sec:decayParameters}}\noindent
While the proposed approach is not overly sensitive to the choice of the algorithmic constants $\depa_1$, $\depa_2$, we must select appropriate values for the residual to decay over a reasonable interval of the parameter $\mu$. Keeping in mind the fundamental idea that the objective function is assumed to decrease as $ \exp(-\depa\mu)$, it is intuitive to choose these constants such that the residual becomes negligible within a small fraction of the considered interval of $\mu$. In the context of waveguide modeling, we can relate this interval to the dimensionless frequency $\dlFreq$, defined as \cite{Gravenkamp2012}
\begin{equation}
  \dlFreq = \frac{\omega h}{c_t}
\end{equation}
with the layer thickness $h$ and the (smallest) shear wave velocity $c_t$. For a multi-layered system, we use the average dimensionless frequency over all layers. A reasonable approach to choosing $\depa_1$ is obtained by requiring that the residual reduces by a factor of $1/e$ within an interval $\Delta \dlFreq$, significantly smaller than the largest dimensionless frequency of interest. This consideration leads, for a single layer, to
\begin{equation}
  \depa_1 = c_{\chi_1}\frac{h^2}{c_t^2}, \quad \depa_2 = c_{\chi_2}\frac{h^2}{c_t^2}
\end{equation}
with constants $c_{\chi_1}, c_{\chi_2}$. 
If the waveguide consists of $n_\ell$ layers with individual thicknesses $h_i$ and $c_{t,i}$, we adapt these equations accordingly:
\begin{equation}\label{eq:depa_inh}
  \depa_1 = \frac{c_{\chi_1}}{n_\ell}\sum \limits_{i=1}^{n_\ell} \frac{h_i^2}{c_{t,i}^2}, \quad
  \depa_2 = \frac{c_{\chi_2}}{n_\ell}\sum \limits_{i=1}^{n_\ell} \frac{h_i^2}{c_{t,i}^2}.
\end{equation}
When computing dispersion curves, we are typically interested in dimensionless frequencies roughly up to $a_{0,\mathrm{max}} = 10$, including the first five to ten propagating modes \cite{Gravenkamp2012}. Hence, we may choose $c_{\chi_1} = 100$, i.e., a decay by a factor of $1/e$ within an interval $\Delta \dlFreq = 0.1$ -- typically about one percent of the entire frequency range of interest.
In our numerical studies, we have found that the algorithm works well if we set $\depa_1 = \depa_2$. However, we required somewhat fewer steps in the ODE solver by choosing a smaller value, e.g., $\depa_2 = \depa_1/10$. This is a plausible choice, as $\depa_2$ affects only the residual of the normalization. While the residual of the eigenvalue problem can take relatively large values whose decay is governed by $\depa_1$, we can enforce the normalization to be exactly satisfied from the start. Hence, we do not require a rapid decrease of the residual in the normalization; in addition, the normalization does not have to be satisfied exactly as long as the error in eigenvalues and eigenvectors remains small. 
Our numerical examples will be performed with $c_{\chi_1} = 100, c_{\chi_2} = 10$, and we will give some numerical evidence for the suitability of this choice at the end of Section~\ref{sec:numex_leaky}. 

\section{Initial values\label{sec:initial}}\noindent
The proposed approach -- like all methods based on mode tracing -- requires adequate starting values, which, in our case, can be viewed as initial values of the system of ordinary differential equations \eqref{eq:leaky_ODE} or \eqref{eq:leaky_ODE_constraint}. Hence, we need to obtain eigenvalues and eigenvectors of the parameter-dependent eigenvalue problem at some value of the parameter (here, the frequency) -- at least approximately. There are numerous ways in which we can attempt to obtain such starting values. The first one is obvious: We can make use of the direct numerical method presented in \cite{Gravenkamp2025}, which allows us to compute all solutions at a given frequency, except that this approach is computationally expensive and, hence, suitable for reasonably small matrix sizes. In this scenario, we would employ the expensive direct solution only once to obtain initial values and trace all required modes starting there. This strategy may still have merit -- particularly when we do not require all modes or we want to follow the behavior of the modes along the frequency -- but the benefit may not always be worth the effort of dealing with these two solution procedures. On the other hand, it is much more exciting to explore the possibilities of \textit{approximating} solutions at a given frequency to obtain starting values. As we demonstrated in Section~\ref{eq:minimalEx}, the mode-tracing algorithm can converge to a branch even when the initial values are very far from a true solution. However, in the case of eigencurves (or any system of coupled equations), the challenge consists in finding \textit{all} solutions. Thus, we need an adequate number of sufficiently good approximate solutions, such that every branch will be reached by one of the starting values. Ideally, we will achieve this by using exactly as many initial values as there are correct solutions, so that we do not waste resources finding the same branch multiple times. 

For our problem at hand, we can attempt to approximate the nonlinear terms that occur in the eigenvalue problem, i.e., approximate the functions of the form $\sqrt{\frac{\omega_0^2}{c_j^2} - k^2}$ at a given frequency $\omega_0$. One particularly simple approximation is obtained as
\begin{equation}
  \sqrt{\frac{\omega_0^2}{c_j^2} - k^2} \approx \frac{\omega_0}{c_j}\qquad\qquad \textit{dashpot approximation},
\end{equation}
which implies that the wavenumber in the half space $\frac{\omega_0}{c_j}$ is significantly larger than that in the waveguide. This is a decent approximation at high frequencies as long as the free-field wave speeds in the waveguide are larger than those in the half space, which is often true in realistic scenarios. In fact, this assumption leads to roughly the same formulation as the dashpot boundary condition that has been presented in \cite{Gravenkamp2014b, Gravenkamp2015}, except that, in the previous work, the interaction with the half space was considered as a von Neumann boundary condition. There, the applicability to many relevant scenarios has been discussed, as well as the limitations of this approach.
To build on this idea, we propose to employ the approximation 
\begin{equation}
  \sqrt{\frac{\omega_0^2}{c_j^2} - k^2} \approx \sqrt{\frac{\omega_0^2}{c_j^2} - \bar{k}^2}\qquad\qquad \textit{mean value approximation}.
\end{equation}
Here, we replace the unknown $k$ by a characteristic wavenumber $\bar{k}$. This value is obtained by assuming that the real part of a mode propagating in the waveguide will typically be between zero and the largest free-field wavenumber of the waveguide's material at the given frequency
\begin{equation}
  \bar{k} = \frac{\omega_0}{2 c_\mathrm{min}},
\end{equation}
where $c_\mathrm{min}$ denotes the minimum wave speed in any of the materials inside the waveguide. Substituting this approximation in each of the nonlinear terms in \eqref{eq:GWproblem_embedded_combined} yields a simple quadratic eigenvalue problem that can be solved straightforwardly, even for large matrices. We note in passing that this is a much better approximation on average\footnote{For most modes, $\bar{k}$ is a better approximation than assuming $k=0$ in the nonlinear terms, especially at relatively large frequencies as all waveguide modes tend towards a free-field solution. However, modes close to their cut-off frequency exhibit a small wavenumber; hence, there will always be a few modes for which this approximation is worse than the simple dashpot.} than the one originally proposed in \cite{Gravenkamp2014b}. As we will see in the numerical examples, this approximation yields excellent starting values that are suitable to trace (almost all) the modes in all examples we have tried. The only exception we have encountered is the quasi-Scholte modes at a fluid/solid interface, which are characterized by $\frac{\omega_0}{c_j^f} \approx k$ and hence are not well represented by any of the approximations above. In many applications, we may simply decide to ignore these modes, as they are often of little practical relevance.\footnote{An important exception can be found in the design of surface acoustic wave (SAW) filters.} Nevertheless, we found that we can simply approximate their wavenumber as  
\begin{equation}
  k \approx \frac{\omega_0}{c^f_j} \qquad\qquad \textit{quasi-Scholte approximation}.
\end{equation}
The corresponding eigenvector is well approximated by setting only the component describing the acoustic pressure in the fluid to one and all other components to zero. In conclusion, we use the mean value approximation and estimate one additional mode for each fluid/solid interface by the quasi-Scholte approximation. This combination was found to yield very reliable initial values such that all eigencurves are found by the mode-tracing algorithm in all examples we have studied. 

\section{Numerical Examples\label{sec:numex}}\noindent
In this section, we demonstrate the applicability of the proposed approach to problems of different complexity. We begin with a minimal example that can easily be reproduced and verified against an analytical solution. In the ensuing, we present four examples of increasing computational demand, involving both homogeneous and layered structures coupled to fluid or solid unbounded media. 

\subsection{Minimal example \label{sec:numex_minimal}}
\begin{figure}\centering
  \subfloat[real part]{\includegraphics[width=0.48\textwidth]{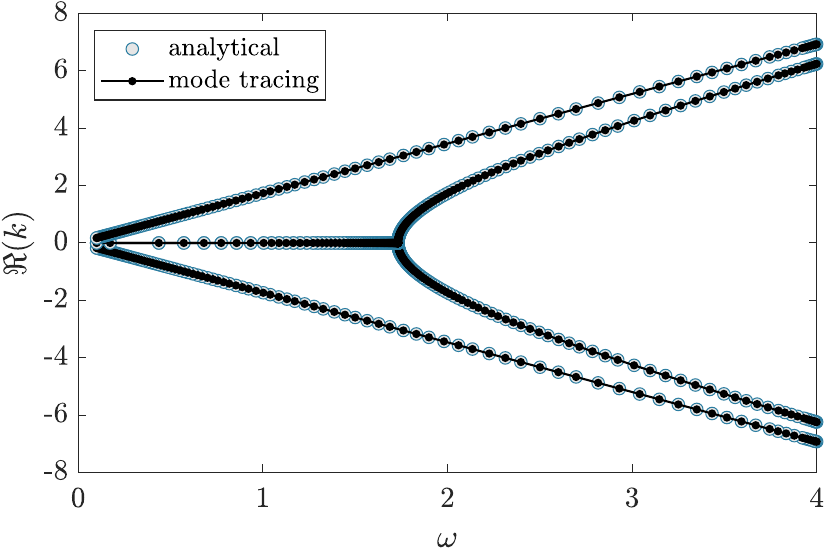}}\hfill
  \subfloat[imaginary part]{\includegraphics[width=0.48\textwidth]{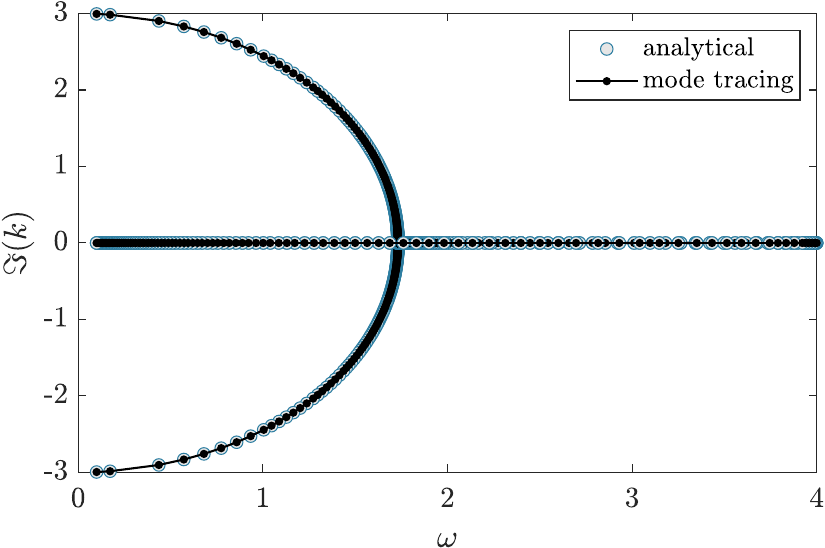}}\\
  \subfloat[error]{\includegraphics[width=0.55\textwidth]{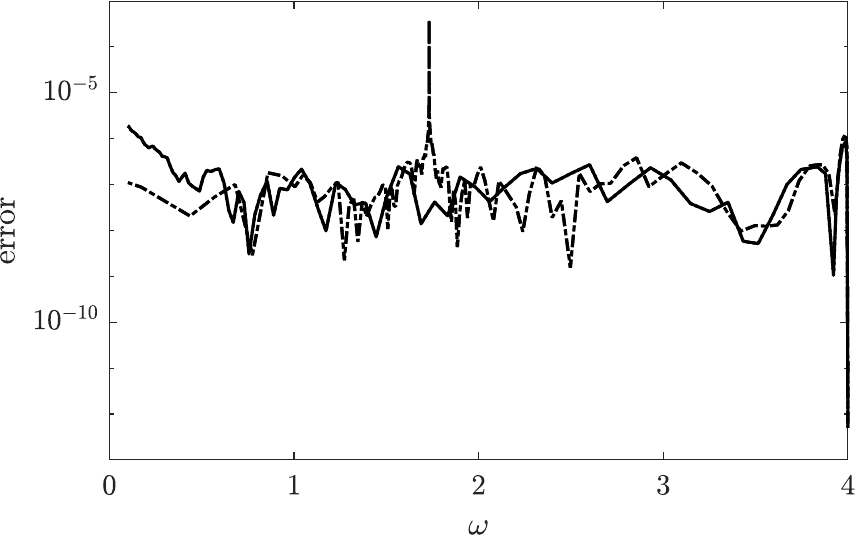}}
  \caption{Minimal example involving matrices of size $2\times 2$. The real (a) and imaginary (b) parts of the eigencurves are compared visually with the analytical solution. The relative error of each eigenvalue within the chosen frequency range is presented in (c).  \label{fig:exampleMinimal} }
\end{figure}

\noindent
We begin with a simple case that we discussed in a different context in \cite{Gravenkamp2022}, namely, an eigenvalue problem of the form \eqref{eq:GWproblem_embedded} with
\begin{equation}
  \*M = \left[\begin{array}{ll}
    2 & 1 \\
    1 & 2
    \end{array}\right], \quad 
    \*E_0=\frac{1}{3}\left[\begin{array}{ll}
    2 & 1 \\
    1 & 2
    \end{array}\right], \quad 
    \*E_1=\mathbf{0}, \quad 
    \*E_2=\frac{3}{2}\left[\begin{array}{rr}
    1 & \shortm 1 \\
    \shortm1 & 1
    \end{array}\right], \quad
    \*R=\mathbf{0}.
\end{equation}
This rather academic example can be obtained by considering a homogeneous plate (thickness $h=2$, mass density $\rho = 3$, shear modulus $G = 1$, Poisson's ratio $\nu = 0.25$, plane strain) with horizontal displacements fixed and vertical displacements approximated by only one linear finite element. The four eigencurves are calculated analytically as
\begin{equation}
  k_{1,2} = \pm\sqrt{3}\omega, \qquad k_{3,4} = \pm\sqrt{3\omega^2-9}.
\end{equation} 
Note that, at $\omega = \sqrt{3}$, the solutions $k_{3,4}$ coincide, and both eigencurves are not continuously differentiable, which becomes a challenge for the adaptive ODE solver employed here. To circumvent this issue, we regularize the eigenvalue problem by modifying the coefficient matrices as
\begin{equation}\label{eq:Edamped}
  \hat{\*E}_i = \*E_i (1 - \I\delta).
\end{equation} 
This is equivalent to changing the shear modulus to $G = 1 - \I\delta$, which can be interpreted as including a small amount of material damping with a constant damping coefficient $\delta$. To obtain numerically stable computations, it is sufficient to use $\delta = 10^{-12}$, which is the value chosen for the presented results.
We calculate initial conditions by solving the eigenvalue problem at $\omega_0 = 4$  and employ the mode-tracing algorithm described in Section \ref{sec:versionOne} with a decay parameter $\depa = 10$ to trace the modes towards $\omega = 0$. The Runge-Kutta-based solver with a variable stepsize implemented in Matlab's function \emph{ode15s} is employed for the solution of the ODE.
Results are presented in Fig.~\ref{fig:exampleMinimal}, showing the real and imaginary parts of each mode $i$, as well as the error defined as
\begin{equation}
  \mathrm{error}_i = \frac{|k_{i,\mathrm{exact}}-k_{i,\mathrm{numerical}}|}{\mathrm{max}(|k_{i,\mathrm{exact}}|)}.
\end{equation}
The subscript $i$ indicates that this error is computed separately for each mode, based on the difference between the numerical and the exact solution at the points resulting from the adaptive stepping procedure. The error is normalized by the maximum absolute value of the eigenvalue of the respective mode within the frequency range of interest. The error is primarily influenced by the relative tolerance chosen in the adaptive Runge-Kutta solver, here $10^{-6}$. Except for a small region around $\omega = \sqrt{3}$, where the error in modes 3 and 4 increases to around $10^{-3}$ due to the mentioned non-differentiability of two eigencurves, it remains below the requested tolerance. Note that, due to the problem's symmetry, the errors of modes 1,2 and those of modes 3,4 are indistinguishable in the figure.

\subsection{Leaky waves \label{sec:numex_leaky}}\noindent
Finally, we employ the proposed approach for the computation of dispersion curves of leaky guided waves.
For better reproducibility and conciseness, we use the same models that have been described in detail in a recent publication \cite{Gravenkamp2025}. Implementations of the approach proposed there for these exact setups can be obtained either separately \cite{Gravenkamp2024b} or within the more general toolbox \textit{SAMWISE} \cite{Gravenkamp2024a}. These test cases had been designed to cover a wide range of different scenarios, and the results are thoroughly verified against other techniques, namely a linearization in the special case of symmetric fluid coupling \cite{Kiefer2019} and the Global Matrix Method implemented in the commercial software \textit{disperse} \cite{Pavlakovic1997} for more complex cases. Hence, we can be brief in discussing these examples, as our focus here is solely on a different numerical approach. For details on the physical behavior and the peculiarities of each setup, we refer to the previous work \cite{Gravenkamp2025}. The four examples are summarized in Table~\ref{tab:examples}. In the third scenario, the plate consists of three layers (titanium-brass-titanium); in all other cases, the plate is a homogeneous layer of either brass or titanium. Each layer has a thickness of 1\,mm and is discretized by one finite element of a polynomial order $p_e$ adequate for the chosen frequency range up to $f_\mathrm{max}$. Details on how to choose the element order for such waveguide models can be found in \cite{Gravenkamp2014,Gravenkamp2014e}.
In each scenario, the plate is coupled to one or two halfspaces at its bottom and/or top surface as described by the column \textit{halfspace}. The examples increase in difficulty from the very common and comparatively straightforward case of a plate immersed in a single fluid to the rather extreme scenario in which a metal plate is coupled to two different solid halfspaces with a small acoustic mismatch at the interfaces.  
The elastic constants of each material are listed in Table~\ref{tab:materialParameters}. Only in the first example, we include a small amount of material damping with $\delta = 0.001$ according to Eq.~\eqref{eq:Edamped} to allow for a smoother transition from leaky to trapped modes.

\begin{table} \centering
  \caption{Summary of the numerical examples discussed in section \ref{sec:numex_leaky}, indicating the materials of each halfspace and layer, the element orders $p_e$, number of degrees of freedom $n_\mathrm{dof}$ (size of finite-element matrices), maximum frequency $f_\mathrm{max}$, maximum attenuation $\eta_\mathrm{max}$, and the CPU times for the proposed approach ('tracing') and the direct solution based on multiparameter eigenvalue problems ('MultiParEig').  \label{tab:examples}}
  \label{tab:examplesOverview}
  \renewcommand{\arraystretch}{1} 
  \begin{tabular}{ccccccccc}\Xhline{2\arrayrulewidth}
        & \multicolumn{2}{c}{material} & & & & & \multicolumn{2}{c}{CPU time} \\
        & halfspace & layer & $p_e$ & $n_\mathrm{dof}$ & $f_\mathrm{max}$ & $\eta_\mathrm{max}$ & tracing & MultiParEig\\
       \Xhline{2\arrayrulewidth}
      I & \makecell{water\\water} & brass & 9  & 22& 4 MHz & 2100 dB/m & 1.1\,s & 3.6\,s\\
      \hline
      II & \makecell{---\\Teflon} & brass & 13 & 45 & 7 MHz & 7000 dB/m & 1.1\,s & 75\,s\\
      \hline
      III & \makecell{Teflon\\oil} & \makecell{titanium \\brass \\ titanium} & \makecell{6 \\8 \\ 6} &  45 & 3 MHz & 2000 dB/m & 1.0\,s & 134\,s\\
      \hline
      IV & \makecell{Teflon\\brass} & titanium & 13 & 32 & 10 MHz & 30000 dB/m & 0.8\,s & 341\,s\\\Xhline{2\arrayrulewidth}
  \end{tabular}
\end{table}

\begin{table} \centering
  \caption{Overview of material parameters used in the numerical experiments. \label{tab:materialParameters}}
  \begin{tabular*}{0.75\textwidth}[tb]{lcccrrr}\Xhline{2\arrayrulewidth}
    & density $\rho$ & \multicolumn{2}{c}{ wave speeds $c_\ell, c_t$ } &  \multicolumn{2}{c}{ Lam\'e parameters $\lambda, G$}\\ \Xhline{2\arrayrulewidth}
    brass   & 8.40 g/cm$^3$ & 4.40 km/s & 2.20 km/s & 81.312 GPa &  40.656 GPa \\
    Teflon  & 2.20 g/cm$^3$ & 1.35 km/s & 0.55 km/s & 2.679 GPa & 0.666 GPa \\
    titanium & 4.46 g/cm$^3$ & 6.06 km/s & 3.23 km/s & 70.726 GPa & 46.531 GPa \\
    water   & 1.00 g/cm$^3$ & 1.48 km/s &         &  \\
    oil     & 0.87 g/cm$^3$ & 1.74 km/s &         & \\ \Xhline{2\arrayrulewidth}
  \end{tabular*}
\end{table}

In all examples, we compute initial conditions at the largest frequency of interest $f_\mathrm{max}$ using the approximation detailed in Section~\ref{sec:initial} and trace each mode towards $f = 0$. Note that the mode tracing is formulated in terms of the squared circular frequency $\mu$, which is converted into the temporal frequency $f$ for the plots only. For conciseness and comparability, we present those modes that are characterized by all partial wave vectors in the halfspaces pointing away from the plate. For a discussion on incoming and radiating waves and methods for their distinction, we refer again to \cite{Gravenkamp2025}. Furthermore, we remove the strongly attenuated (\textit{non-propagating}) modes, i.e., solutions with an attenuation above the chosen value of $\eta_\mathrm{max}$ listed in Table~\ref{tab:examples}.

\begin{figure}\centering
  \subfloat{\includegraphics[width=0.5\textwidth]{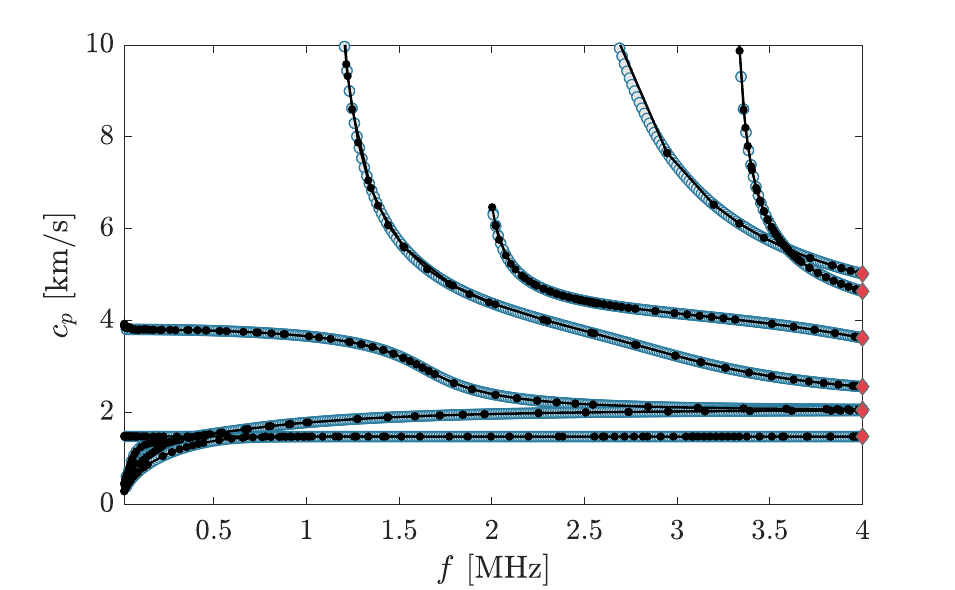}}\hfill
  \subfloat{\includegraphics[width=0.5\textwidth]{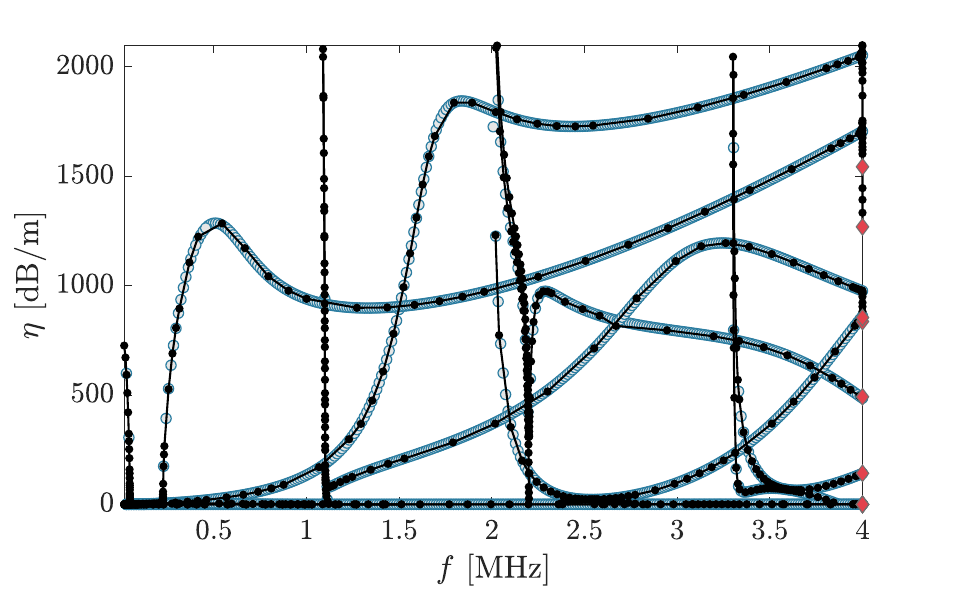}}\\
  \subfloat{\includegraphics[width=0.5\textwidth]{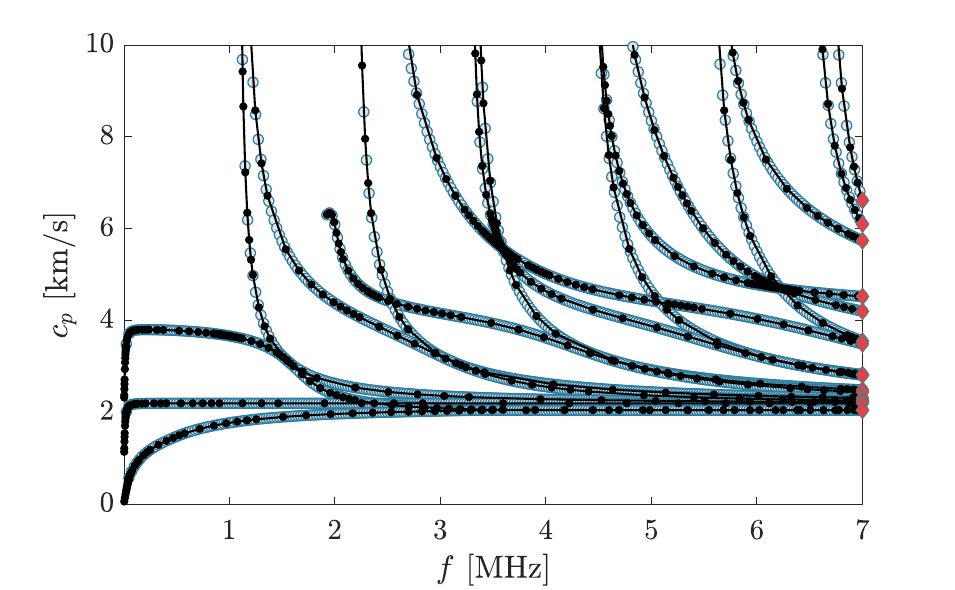}} \hfill
  \subfloat{\includegraphics[width=0.5\textwidth]{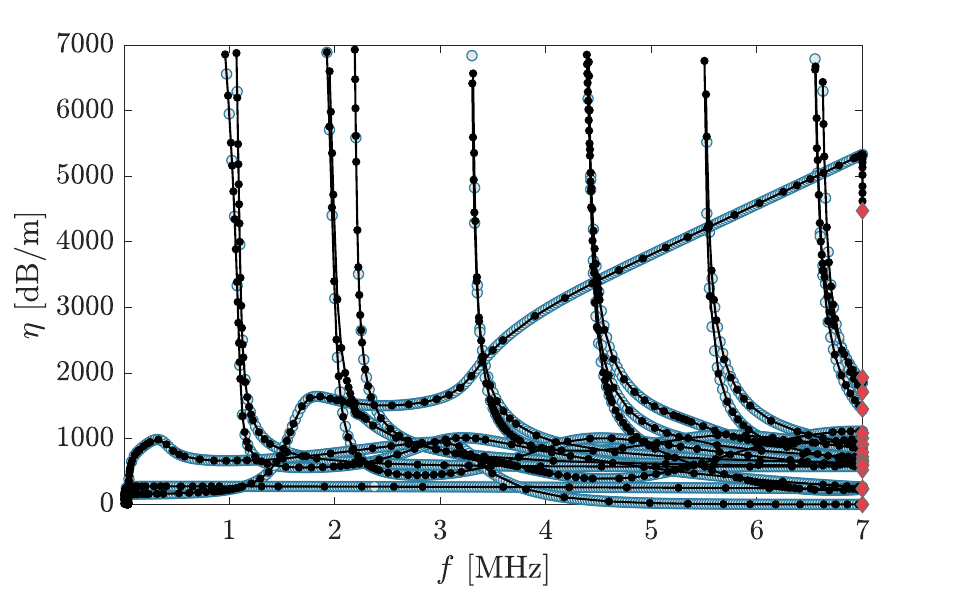}}\\
  \caption{Dispersion curves computed using the proposed mode tracing approach ($\mathbf{-}\negthickspace\negthickspace\bullet\negthickspace\negthickspace\mathbf{-}$) and direct computation (\tikzcircle[mybluelight, fill=mygraylight]{1.5pt}) for examples I (top) and II (bottom). \label{fig:examples12} }
\end{figure}

\begin{figure}\centering
  \subfloat{\includegraphics[width=0.5\textwidth]{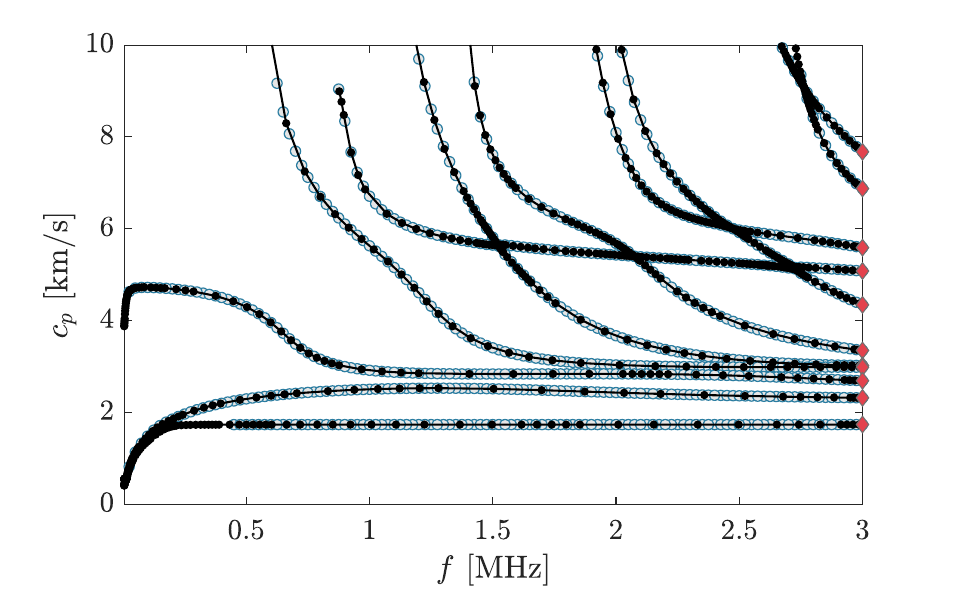}}\hfill
  \subfloat{\includegraphics[width=0.5\textwidth]{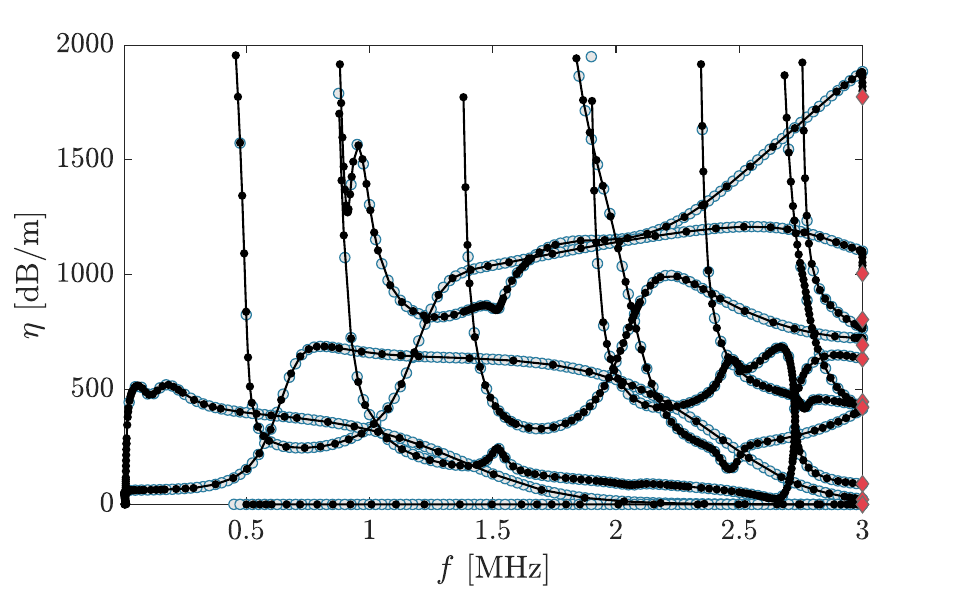}}\\
  \subfloat{\includegraphics[width=0.5\textwidth]{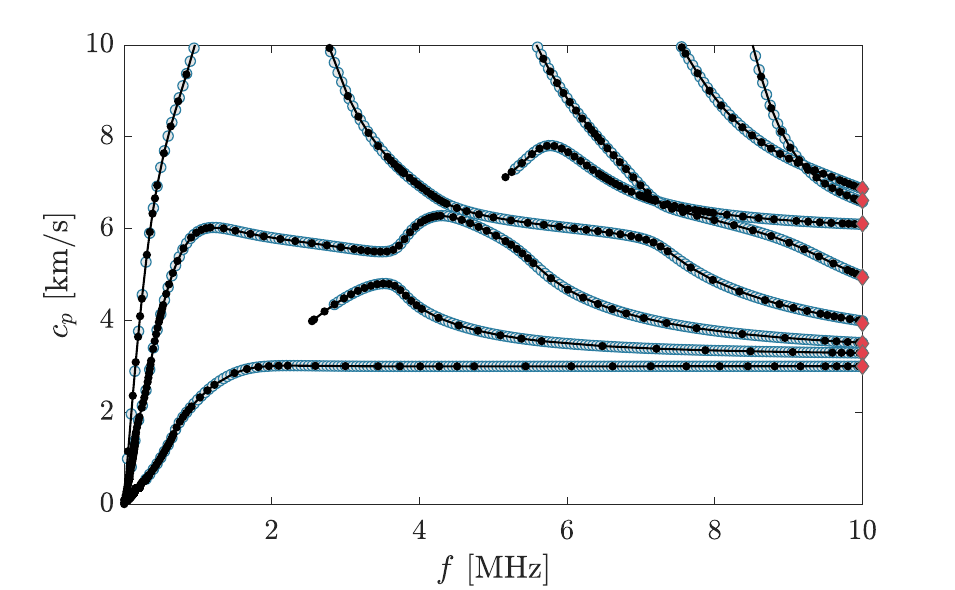}}\hfill
  \subfloat{\includegraphics[width=0.5\textwidth]{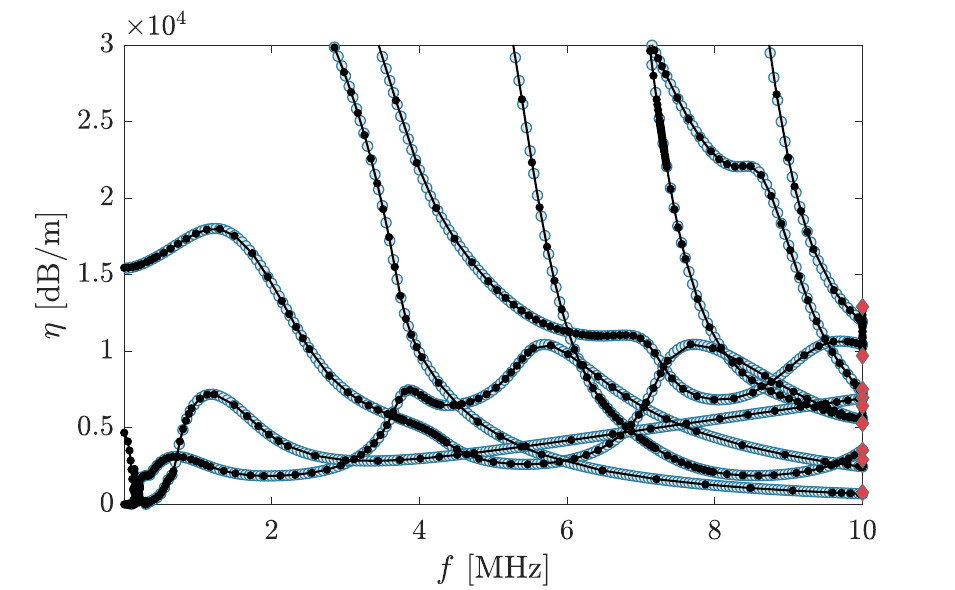}}
  
  \caption{Dispersion curves computed using the proposed mode tracing approach ($\mathbf{-}\negthickspace\negthickspace\bullet\negthickspace\negthickspace\mathbf{-}$) and direct computation (\tikzcircle[mybluelight, fill=mygraylight]{1.5pt}) for examples III (top) and IV (bottom). \label{fig:examples34} }
\end{figure}

The dispersion curves in terms of phase velocities and attenuation are presented in Fig.~\ref{fig:examples12} for examples I, II and in Fig.~\ref{fig:examples34} for examples III, IV. The results of the proposed approach are compared with those obtained in \cite{Gravenkamp2025} by direct solution of the multiparameter eigenvalue problem. In all cases, both computations are in excellent agreement.  Note that the approximation described in Section~\ref{sec:initial} generally yields very good initial values at high frequencies, with discrepancies mainly visible in the attenuation of strongly attenuated modes. However, even in the cases where the initial values are relatively poor approximations (see the attenuation in example I), the mode tracing reaches highly accurate solutions within a few steps. We may also highlight the fact that the quasi-Scholte modes have successfully been computed in both examples I and III. Another interesting detail is that the mode-tracing approach is capable of following eigencurves that exhibit rapid changes, which could be missed by a constant frequency step; see, e.g., the almost vertical lines in the attenuation of example I, Fig.~\ref{fig:examples12}.

Table~\ref{tab:examples} also lists the computational times required by both approaches for the specific examples studied here. Care must be taken when drawing general conclusions, as the computational costs depend on many factors. Obviously, the costs increase with the size of the finite-element matrices, and, especially in the case of the multiparameter eigenvalue problem, they depend drastically on the number of partial waves in the unbounded media. Furthermore, when employing the mode-tracing algorithm, the frequency steps are adjusted automatically based on the requested relative tolerance of the ODE solver (here 0.01), while we choose a fixed frequency resolution for the 'MultiParEig approach.' Another important difference lies in the fact that, in the mode-tracing algorithm, we can choose to compute only specific modes (in the numerical examples, those are the lowly attenuated modes with wave vectors of partial waves in the unbounded domains pointing away from the plate). Perhaps most importantly, the implementation of the mode-tracing algorithm is not optimized for efficiency, as it involves calls to Matlab functions at every iteration in every step for every mode. In comparison, the bottleneck of the MultiParEig approach is a highly optimized eigenvalue problem solver. Despite all these difficulties in comparing the different techniques, the data serves to give the reader a rough idea of the efficiency of the mode-tracing approach, with CPU times in the order of one second for reasonable examples. 

\begin{figure}\centering
  \includegraphics[height=0.28\textheight]{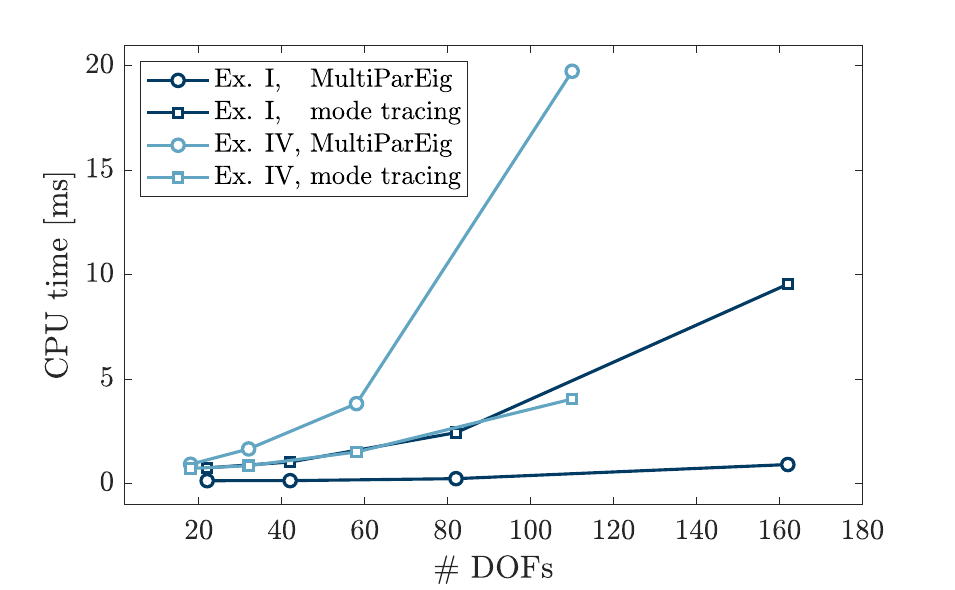}
  \caption{Computational times per mode and frequency in examples I and IV. The number of degrees of freedom ($\#$ DOFs) is varied by increasing the element order $p_e$. Results are computed using the mode-tracing approach, as well as the formulation based on multiparameter eigenvalue problems (MultiParEig). \label{fig:cpuTimes} }
\end{figure}

To obtain a different perspective, we evaluate the CPU times per frequency and mode, i.e., the average cost of obtaining one individual solution in the dispersion diagram. This is mainly relevant if we wish to compute all modes, including the evanescent ones. To this end, we used examples I and IV (the simplest and the most complex) and computed the solution for varying matrix sizes by increasing the polynomial degree of the finite-element approximation. We then divided the total computational time by the number of computed modes and the number of frequency steps, leading to an average CPU time per solution. This value is plotted in Fig.~\ref{fig:cpuTimes} against the number of degrees of freedom for both examples and both approaches. We can see that the MultiParEig approach is highly efficient in the simple case of Example I, whereas, in the more complex Example IV, computational costs rapidly increase with the matrix size, making the mode tracing much more effective, even when computing all solutions.

\begin{figure}\centering
  \subfloat{\includegraphics[width=0.5\textwidth]{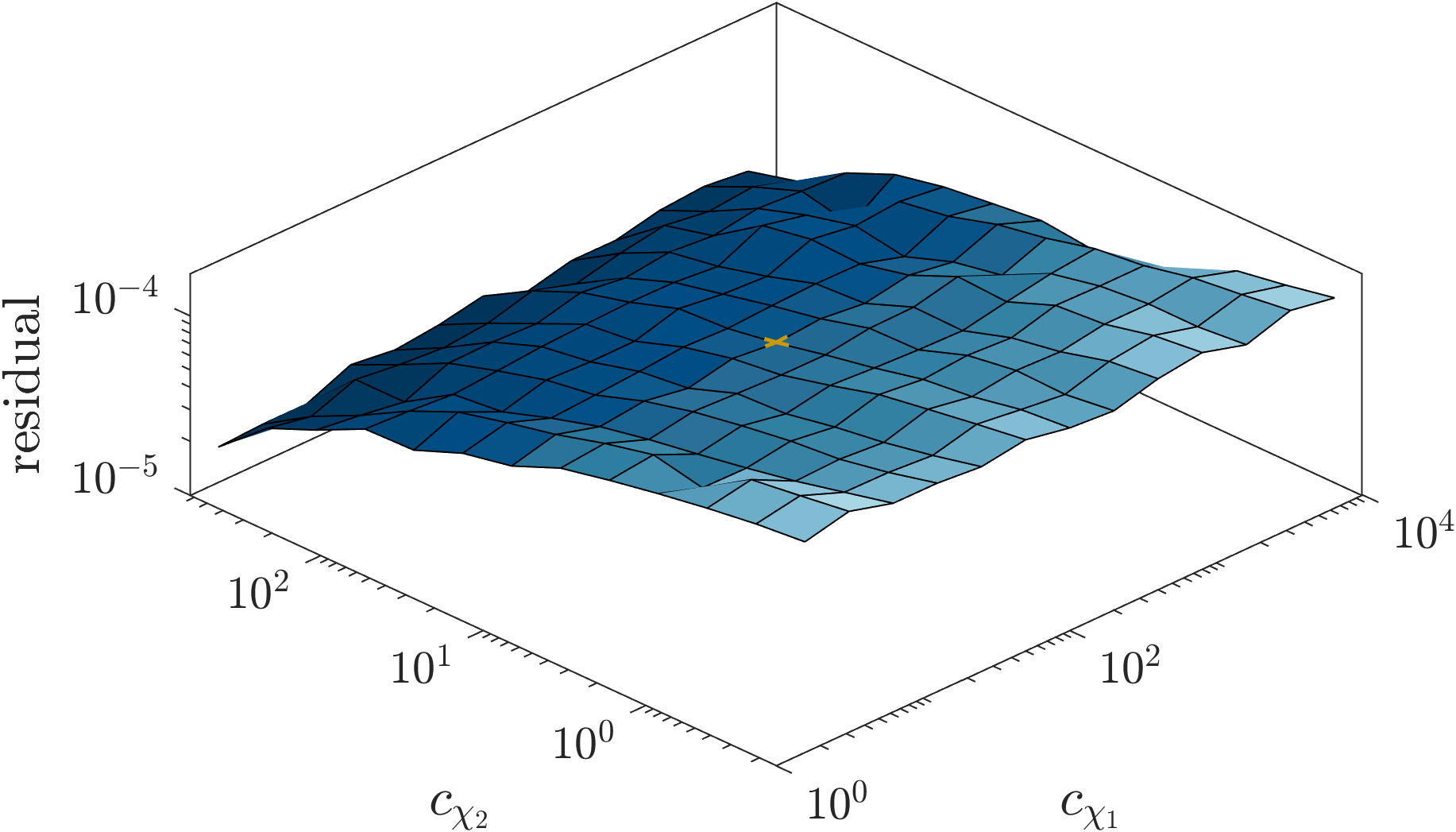}}\hfill
  \subfloat{\includegraphics[width=0.5\textwidth]{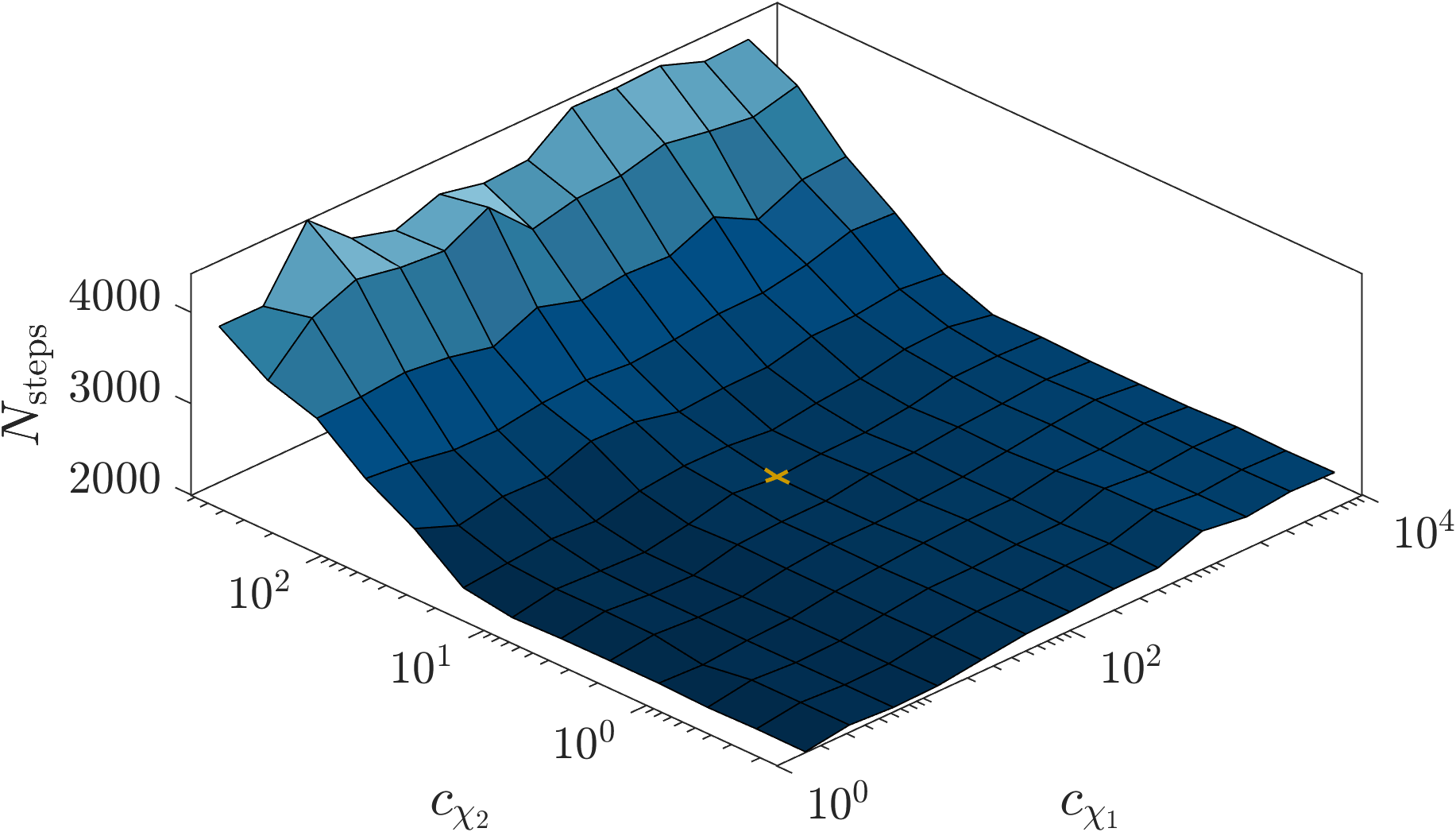}}
  
  \caption{Average residual and total number of steps taken by the ODE solver when varying the decay parameters $c_{\chi_1}$ and $c_{\chi_2}$ in example III. \label{fig:etaStudy}}
\end{figure}

Finally, we studied the effect of varying the decay parameters -- specifically $c_{\depa_1}$, $c_{\depa_2}$ according to the definition in Eqs.~\eqref{eq:depa_inh} -- on the solution's accuracy and efficiency. To assess the former, we evaluated the residual $r_i$ of the objective function \eqref{eq:evp_objective} for each individual solution $(\egv_i,k_i,\mu_i)$, normalized in the following way
\begin{equation}
  r_i = |\*r_i|, \qquad
  \*r_i =
   \begin{bmatrix}
    \frac{\Mflow(k_i,\mu_i)\,\egv_i}{||\Mflow(k_i,\mu_i)||_\mathrm{F}} \\
    \norma(\egv_i)
  \end{bmatrix},
\end{equation}
where $||\dotP||_\mathrm{F}$ denotes the Frobenius norm and $|\dotP|$ the Eucledian vector norm. This residual was then averaged over all solutions. For this computation, we decreased the relative tolerance of the ODE-solver to $10^{-4}$ to isolate the effect of the decay parameter from the accuracy in solving the ODE. In addition, we took note of the number of steps $N_\text{steps}$ required by the solver. For brevity, we present in Fig.~\ref{fig:etaStudy} these values only for example III, as it involves a layered plate and coupling to both a fluid and a solid halfspace; however, results for the other examples are similar. First of all, we can observe that the proposed method is remarkably robust for a wide range of decay parameters (note the logarithmic scales of $c_{\depa_1}$, $c_{\depa_2}$). There is, however, a trade-off between a low residual and a small number of steps. This is quite intuitive, as a small decay parameter results in a slow decay of deviations from the exact solution and hence to a larger average residual. On the other hand, a too large decay parameter forces rapid changes in the solution due to perturbations, which can require a finer resolution, hence a larger number of steps. We may also note that both criteria are less sensitive to changes in $c_{\depa_1}$ compared to $c_{\depa_2}$. Nevertheless, this study confirms that the choice made for the numerical examples, namely, $c_{\depa_1}=100$, $c_{\depa_2} = 10$ (marked by crosses in Fig.~\ref{fig:etaStudy}), is indeed a very suitable trade-off.

\section{Conclusions}\label{sec:conclusion}\noindent
We have seen that the parameter-dependent nonlinear eigenvalue problems arising in waveguide models can be solved in a rather unconventional way by transformation into a system of ordinary differential equations and tracing each mode along the frequency axis starting from some initial value. The advantages are that this method involves only the original finite-element matrices of the waveguide problem (in contrast to the operator determinants assembled in the solution of the multi-parameter eigenvalue problem), can exploit standard ODE solvers, and allows the computation of individual modes. Perhaps most importantly, the method is remarkably robust even when only approximate solutions are available as starting values. This property makes the application of this idea worth exploring for various other scenarios where a nonlinear eigenvalue problem is challenging to solve, but reasonable approximations exist.

\section*{Acknowledgments}\noindent
Bor Plestenjak has been supported by the Slovenian Research and Innovation Agency (grant P1-0294). Daniel A.\ Kiefer has received support under the program ``Investissements d'Avenir'' launched by the French Government under Reference No.\ ANR-10-LABX-24.

\appendix

\bibliographystyle{elsarticle-num}
\bibliography{embeddedWaveguide.bib,embeddedTracing.bib,references_coauthors.bib}

\end{document}

%% file: myColorsMatlab.tex
\definecolor{myred}{rgb}{0.6353,0.0784, 0.1843}
\definecolor{myblue}{rgb}{0, 0.2588, 0.6275}
\definecolor{mygreen}{rgb}{0, 0.5333, 0.2400}
\definecolor{mypurple}{rgb}{0.5882, 0.1176, 0.7059}
\definecolor{myorange}{rgb}{0.8431, 0.3333, 0.0980}
\definecolor{mygray}{rgb}{0.5020, 0.5020, 0.5020}
\definecolor{mygraylight}{rgb}{0.9, 0.9, 0.9}
\definecolor{mybluelight}{rgb}{0.1725, 0.4902, 0.6275}
\definecolor{mybrown}{rgb}{0.6275, 0.3216, 0.1765}

%% file: plate_geometry_embedded.tex

\begin{tikzpicture}[>=stealth,
plate/.style={fill=black!10, minimum width=\l, minimum height=\h, inner sep=0pt},
]
\footnotesize
\pgfmathsetmacro\h{1 cm}
\pgfmathsetmacro\l{5 cm}
\pgfmathsetmacro\kxlen{1.5 cm}

\node[plate] (plate) at (0, 0) {};
\draw[thick, dashed] (plate.north west) -- (plate.north east); 
\draw[thick,dashed] (plate.south west) -- (plate.south east); 

\draw[Bar-Bar] ([xshift=4pt]plate.north east) -- node[right]{$h$} ([xshift=4pt]plate.south east);

\coordinate (O) at ([xshift=-1cm, yshift=-0.5em]plate.center);
\coordinate (Pkx) at ([xshift=\kxlen]O);
\draw[->, thick] (O) -- node[above, inner sep=2pt, yshift=3pt]{quasi-guided wave} (Pkx) node[right, inner sep=2pt]{$k$};

\node[right, yshift=0pt, xshift=-30pt] at (plate.east) {$\ten{C}, \rho$};
\node[mybrown, above right, yshift=3pt, xshift=-30pt] at (plate.north east) {$\widetilde{\ten{C}}_1, \tilde{\rho}_1$};
\node[mybluelight, below right, yshift=-3pt, xshift=-30pt] at (plate.south east) {$\tilde{c}_2, \tilde{\rho}_2$};

\node[mybrown, above right, yshift=3pt] at (plate.north west) {solid};
\node[mybluelight, below right, yshift=-3pt] at (plate.south west) {fluid};

\begin{scope}[shift={(-0.5*5cm, 0)}]
    \draw[-latex, fill opacity = 1] (0, 0) -- (1.5em, 0) node[above, inner sep=2pt]{$\dirvec{x}$};
    \draw[-latex, fill opacity = 1] (0, 0) -- (0, 1.5em) node[left, inner sep=2pt]{$\dirvec{y}$};
    \draw[fill=white] (0,0) circle (3pt) node[below left, inner sep=4pt]{$\dirvec{z}$};
    \draw[fill=black] (0,0) circle (1pt);
\end{scope}

\coordinate (Oupp) at ($(O)+(0,0.5cm+0.5em+3pt)$);
\draw[mybrown, ->, thick]  (Oupp) -- node[pos=0.6,below right,outer sep=0pt,inner sep=1pt]{$\kappa_1$} +(1.5cm,0.6cm);
\draw[mybrown, ->, thick]  (Oupp) -- node[above left,outer sep=0pt,inner sep=1pt]{$\gamma_1$} +(1.5cm,1cm);

\coordinate (Obot) at ($(O)-(0,0.5cm-0.5em+3pt)$);
\draw[mybluelight, ->, thick]  (Obot) -- node[pos=0.6,above right,outer sep=0pt,inner sep=1pt]{$\kappa_2$} +(1.5cm,-0.5cm);

\end{tikzpicture}